\documentclass[10pt,a4paper]{article}

\usepackage[utf8]{inputenc}
\usepackage[T1]{fontenc}
\usepackage{mathpazo}  
\usepackage[margin=2.5cm]{geometry}
\usepackage{graphicx}
\usepackage{amsmath,amssymb}
\usepackage{booktabs}
\usepackage{tabularx}
\usepackage{multirow}
\usepackage{float}
\usepackage{caption}
\usepackage[super,sort&compress]{natbib}
\usepackage{hyperref}
\usepackage{xcolor}
\usepackage{lineno}
\usepackage{setspace}
\usepackage{microtype}
\usepackage{textcomp}

\hypersetup{colorlinks=true, linkcolor=blue!60!black, citecolor=blue!60!black, urlcolor=blue!60!black, pdfsubject={v20-reviewer-fixes-20260311}}
\title{\Large\bfseries Battery health reporting fails independent validation across manufacturers}

\author{
  Jeongju Park$^{1}$,
  Kyungkak Kim$^{2}$,
  Seungho Geum$^{2}$,
  Junhyung Lee$^{1}$,
  Hyeongyu Son$^{1}$,\\
  Sekyung Han$^{1,2,*}$
  \\[6pt]
  {\normalsize $^1$Department of Electrical Engineering, Kyungpook National University,}\\
  {\normalsize Daegu, Republic of Korea}\\
  {\normalsize $^2$Betterwhy Inc., 24-13 2Sandan2-gil, Waegwan-eup, Chilgok-gun,}\\
  {\normalsize Gyeongbuk, Republic of Korea}\\[6pt]
  {\normalsize $^*$Correspondence: skhan@knu.ac.kr}
}

\date{March 2026}

\begin{document}

\maketitle
\thispagestyle{empty}

\noindent
\textbf{Battery state-of-health (SOH) reported by on-board battery management systems (BMS) is the primary metric available to electric vehicle (EV) owners and regulators, yet no study has validated its reliability across manufacturers against independent measurements. Here we show, through an epidemiological study of 1,114~EVs spanning five manufacturers and 375~days, that battery health reporting is fundamentally unreliable: real capacity differences of 12--25\% exist within every model, but BMS~SOH fails to track them, with correlations ranging from $\rho = 0.10$ (non-significant) to $\rho = 0.62$ only under restrictive filtering, while 384~vehicles do not expose SOH at all. A manufacturer-independent electrochemical marker achieves 74--89\% degradation classification accuracy across all platforms without requiring BMS data, and a controlled laboratory validation on cells identical to those in the fleet confirms that partial-voltage-window charge measurements track reference capacity with $\rho > 0.80$ across all 60 voltage windows ($\textit{p} < 0.001$). These findings reveal a structural information asymmetry with direct implications for the EU Battery Regulation's 2027 SOH transparency mandate, California's Advanced Clean Cars (ACC)~II durability requirements, warranty enforcement, used-vehicle valuation, right-to-repair legislation, and second-life battery markets.}

\vspace{1em}

The global electric vehicle (EV) fleet has grown rapidly, with over 40~million battery EVs on roads worldwide as of 2024 and annual sales exceeding 17~million units\cite{IEA2025,BloombergNEF2025}.
As lithium-ion battery technology matures and costs decline\cite{Schmuch2018}, a central concern for EV owners, fleet operators, and the used-car market is battery degradation---the gradual loss of capacity and power capability that directly affects vehicle range and resale value\cite{Wassiliadis2022}.
Battery state-of-health (SOH), typically reported as a percentage of original capacity, is the primary metric used by consumers and regulators to assess battery condition\cite{Tian2020}.

However, SOH as reported by on-board battery management systems (BMS) relies on proprietary algorithms that vary across manufacturers and are not publicly documented\cite{Berecibar2016,Li2019}.
The lack of standardization means that ``SOH\,=\,95\%'' from one manufacturer may represent a fundamentally different battery condition than the same value from another.

This opacity creates a critical information asymmetry---directly analogous to Akerlof's ``market for lemons''\cite{Akerlof1970}---where the entity responsible for warranty liability is simultaneously the sole arbiter of when that liability is triggered.
Battery warranty claims, which commit manufacturers to replace or repair batteries that fall below a specified SOH threshold (typically 70--80\%)\cite{EU2023,CARB2022}, depend entirely on the manufacturer's own SOH calculation.
The emerging second-life battery market further exacerbates this problem, as repurposing decisions depend critically on accurate health assessment\cite{Zhu2021,Martinez2018}.

Laboratory studies have characterized degradation mechanisms including lithium inventory loss, solid-electrolyte interphase (SEI) growth, and cathode structural changes\cite{Birkl2017,Dubarry2012}, while production-induced cell-to-cell variation\cite{Baumhofer2014} and non-linear aging ``knees''\cite{Attia2022} further complicate health assessment.
Fleet-scale analyses have reported aggregate trends using BMS~SOH at face value\cite{Geotab2024,Geotab2025,Recurrent2024}, and data-driven approaches have demonstrated cycle life prediction\cite{Severson2019,Attia2020} and SOH estimation from laboratory data\cite{Ng2020,Roman2021,Wei2025}.
However, no study has quantified cross-manufacturer SOH reliability through a systematic audit using a single, independent measurement protocol applied to a large real-world fleet.
This question has added urgency: the EU Battery Regulation\cite{EU2023} mandates SOH transparency from 2027, and California's Advanced Clean Cars (ACC)~II program\cite{CARB2022} requires minimum battery durability guarantees---yet both frameworks implicitly assume that reported SOH is meaningful, an assumption our data challenge directly.
Our study fills this gap by providing the first independent, cross-manufacturer validation of BMS~SOH against a manufacturer-independent benchmark in an operational fleet of over 1,100~vehicles.

Here we present a cross-manufacturer epidemiological study of 1,114~EVs from Hyundai, Kia, Genesis, Audi, and Volkswagen using continuous telematics data spanning approximately 12~months (375~days).
We develop a BMS-independent capacity measurement protocol based on constant-current (CC) charging segments and validate its consistency across voltage windows.
By comparing these independent measurements with BMS-reported SOH, we reveal that SOH reliability is profoundly model-dependent---ranging from complete non-availability to moderate correlation under controlled conditions---and validate universal, BMS-independent health indicators that work across all platforms.
Our principal contributions are threefold: (1)~the first independent, cross-manufacturer audit of BMS~SOH reliability against a single measurement protocol; (2)~quantification of the resulting information asymmetry and its implications for warranty and resale markets; and (3)~field validation---across 1,114~vehicles and five battery architectures---that established laboratory markers and usage-pattern analysis can serve as manufacturer-independent health indicators without requiring BMS data.

\section*{Results}

\subsection*{Capacity heterogeneity within models}

To establish that real capacity differences exist between vehicles of the same model, we identified CC slow charging sessions from the telematics data and measured the accumulated charge (dQ) across standardized voltage windows (Fig.~\ref{fig:methodology}b)---an approach grounded in differential capacity analysis principles\cite{Bloom2005,Weng2014}.
Sessions were selected using physics-validated BMS charging state flags (Precision\,=\,88\%, Recall\,=\,91\% against a flag-free physics-only detector; see Methods).
We imposed strict controls: current coefficient of variation (CV) $< 25\%$, temperature range 10--35\textdegree C, edge trimming of 3~minutes to exclude transients, and voltage window boundaries aligned to measurement quantization steps.

Pack voltage measurement resolution varies across platforms: E-GMP~192S (97.5~mV pack, 0.51~mV/cell), E-GMP~180S ($\sim$97.5~mV pack, 0.54~mV/cell), Commercial~90S (102.5~mV, 1.14~mV/cell), Niro/Kona~98S (102.5~mV, 1.05~mV/cell), and MEB~96S (252.5~mV, 2.63~mV/cell).
Such controlled conditions are essential because charging behavior strongly influences degradation rate estimates\cite{Pelletier2017,DePalma2023}.

\begin{figure}[H]
\centering
\includegraphics[width=\textwidth]{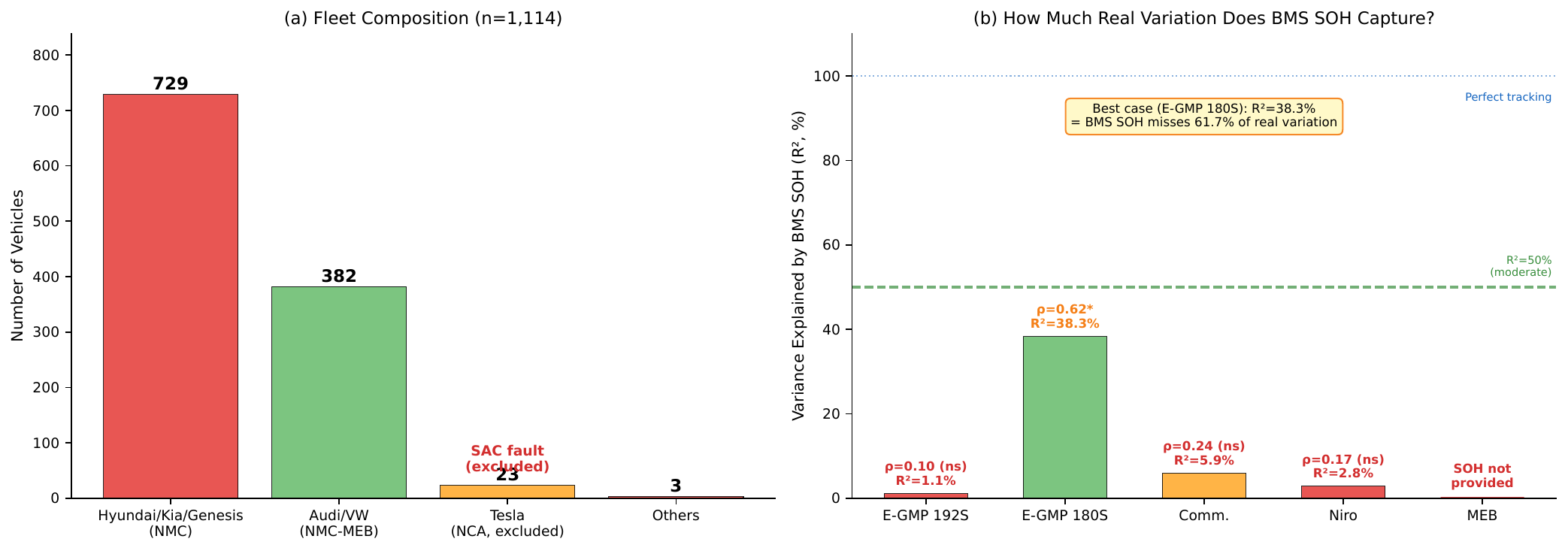}
\caption{\textbf{Dataset overview and SOH landscape across 1,114 electric vehicles.}
\textbf{a,} Fleet composition across five manufacturers, showing vehicle counts and BMS~SOH availability.
\textbf{b,} Variance in independent dQ capacity explained ($R^2$) by BMS~SOH, shown per platform. The best-performing platform (E-GMP~180S) explains 38.3\% of capacity variation; most platforms show near-zero explanatory power, and MEB does not expose SOH through OBD-II.}
\label{fig:dataset}
\end{figure}

Under these controlled conditions, significant inter-vehicle capacity heterogeneity emerged across all model platforms (Fig.~\ref{fig:dataset}a,b; Table~\ref{tab:capacity}).
The Electric-Global Modular Platform (E-GMP) comprises two cell configurations in our dataset: the 192-series (E-GMP~192S; $n = 59$ with CC sessions) showed a 16.4\% dQ gap between the highest and lowest usage quartiles ($\textit{p} = 0.0002$).
Commercial vehicles (Porter/Bongo, $n = 52$) exhibited the largest gap at 24.7\% ($\textit{p} = 0.0003$).
Even the narrowest-range platform, E-GMP~180S ($n = 36$), showed a 12.2\% gap ($\textit{p} = 0.014$).
The Modular Electric Drive (MEB) platform (Audi Q4 / VW ID.4, $n = 152$) displayed a dQ coefficient of variation of 10.3\%.

\subsection*{Methodological validation}

We validated the dQ methodology through three independent approaches.

\begin{figure}[H]
\centering
\includegraphics[width=\textwidth]{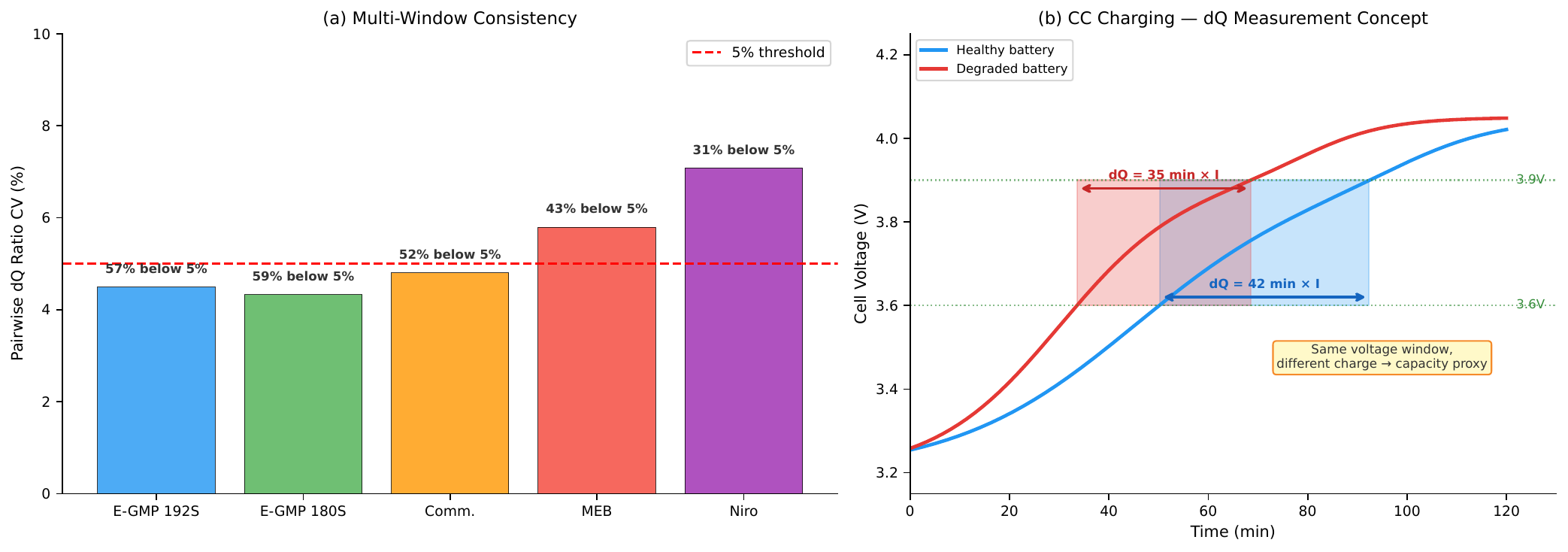}
\caption{\textbf{Validation of the dQ capacity measurement methodology.}
\textbf{a,} Multi-window consistency: distribution of pairwise dQ ratio CV across four non-overlapping 0.1~V/cell voltage windows. Median CV ranges from 4.3\% (E-GMP~180S) to 7.1\% (Niro/Kona); see text for ground truth validation results (Extended Data Table~\ref{tab:groundtruth}).
\textbf{b,} Schematic of the constant-current charging dQ measurement protocol.}
\label{fig:methodology}
\end{figure}

\textbf{Multi-window consistency} (Fig.~\ref{fig:methodology}a).
For each vehicle pair, we computed the dQ ratio across four non-overlapping 0.1~V/cell voltage windows (W1: 3.50--3.60~V, W2: 3.60--3.70~V, W3: 3.70--3.80~V, W4: 3.80--3.90~V).
If dQ reflects true capacity, the ratio between any two vehicles should remain constant regardless of the voltage window examined.
The median pairwise ratio CV was 4.5\% for E-GMP (57\% of pairs below 5\%), 4.3\% for E-GMP~180S (59\% below 5\%), 4.8\% for Commercial (52\% below 5\%), and 7.1\% for Niro/Kona (31\% below 5\%), confirming that dQ differences represent genuine capacity differences rather than measurement artifacts (Fig.~\ref{fig:methodology}a).
The 5\% threshold corresponds approximately to the intra-vehicle measurement CV (5--7\%), below which pairwise ratio variation is attributable to measurement noise rather than methodological inconsistency.
The MEB platform showed slightly higher variability (median CV\,=\,5.8\%, 43\% below 5\%), which we attribute to the wider measurement quantization of its pack voltage sensor (252.5~mV vs.\ 97.5--102.5~mV for Korean platforms).

\textbf{Ground truth validation} (Extended Data Table~\ref{tab:groundtruth}).
To establish external validity, we identified vehicles with wide-range CC charging sessions spanning $\geq 0.25$~V/cell---in-field analogues of laboratory capacity tests---and compared their rankings with those obtained from our standard narrow (0.10~V/cell) windows.
For the E-GMP, the narrow-window dQ ranking matched the wide-window ranking with $\rho = 0.91$ ($\textit{p} < 0.001$, $n = 27$), and Commercial vehicles showed $\rho = 0.75$ ($\textit{p} < 0.001$, $n = 24$).
The Niro/Kona platform showed $\rho = 0.52$ for rank correlation but Pearson $r = 0.96$ (log-transformed), with one outlier vehicle correctly identified as having approximately 55\% of normal capacity---demonstrating that the narrow-window method accurately captures even extreme degradation cases.
E-GMP~180S showed non-significant correlation ($\rho = 0.38$, $\textit{p} = 0.28$, $n = 10$), but this reflects minimal inter-vehicle capacity variation (dQ CV\,=\,1.4\%) rather than methodological failure---when true differences are negligible, rank correlation is inherently unstable.

\textbf{Lab cell validation} (Extended Data Fig.~4).
To confirm that the dQ proxy faithfully tracks ground-truth capacity under controlled conditions, we tested the methodology on accelerated aging data from a 55.6~Ah NCM pouch cell identical to those used in the E-GMP~180S fleet vehicles (see Methods).
Over 198 charge--discharge cycles spanning SOH degradation from 100\% to 9.7\%, the Spearman rank correlation between partial-window dQ and full reference performance tests (RPT) capacity exceeded $\rho = 0.80$ across all 60 tested voltage windows ($\textit{p} < 0.001$; Extended Data Fig.~4b).
The best-performing window (3.4--4.1~V) achieved $\rho = 0.94$, while the best ratio accuracy was obtained at 3.5--4.2~V (MAE\,=\,0.017).
Critically, this validation uses the same cell chemistry and format deployed in the IONIQ~5 vehicles in our fleet, providing a direct bridge between controlled laboratory measurements and the field-derived dQ proxy.

\subsection*{BMS SOH fails across platforms}

\begin{figure}[H]
\centering
\includegraphics[width=\textwidth]{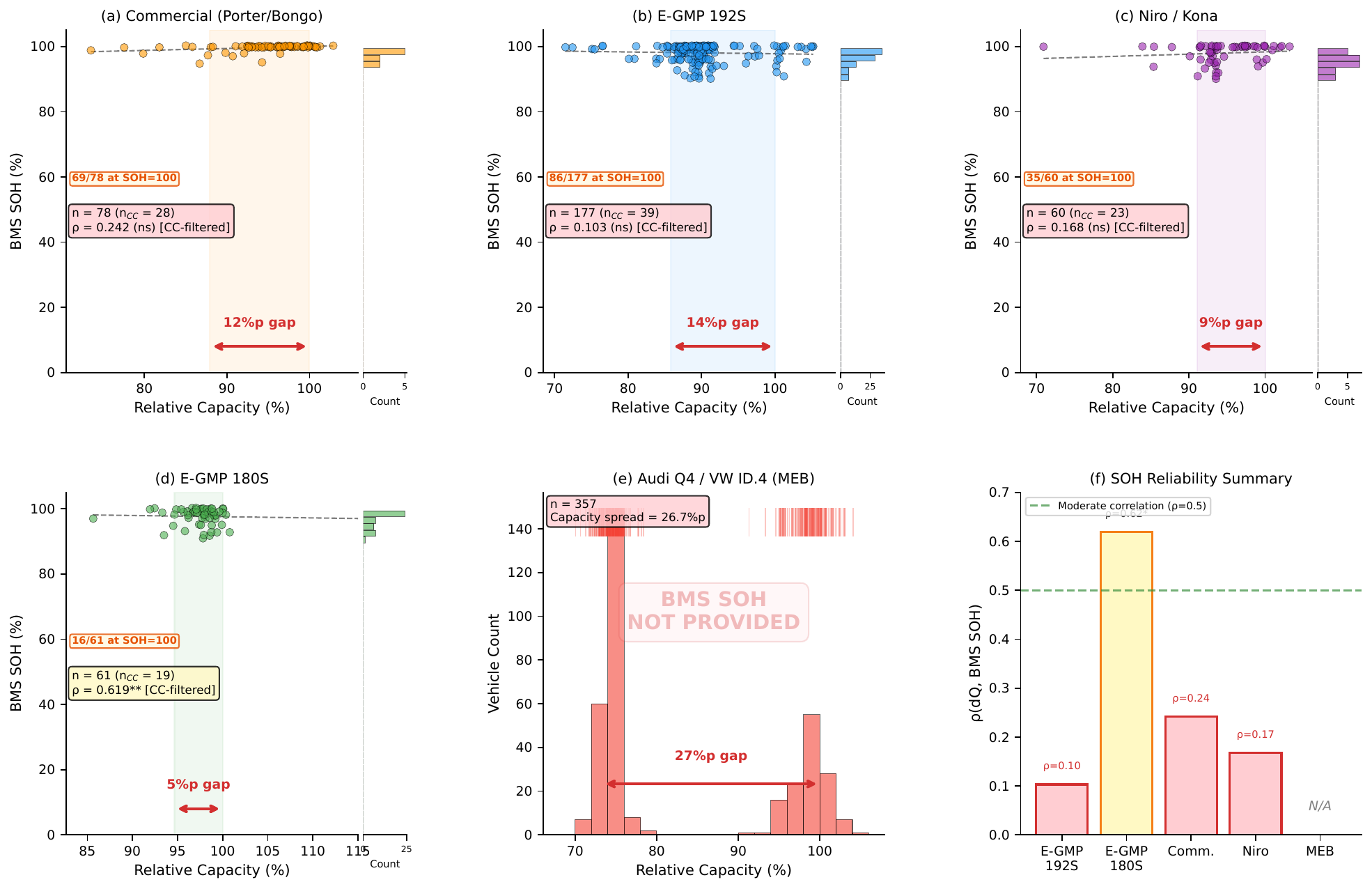}
\caption{\textbf{Model-dependent unreliability of BMS~SOH.}
\textbf{a--d,} Scatter plots of independently measured relative capacity vs.\ BMS~SOH for four platforms with SOH data. All data points are shown ($n$), but Spearman correlations are computed from the CC-filtered subset ($n_{\mathrm{CC}}$; Table~\ref{tab:capacity}).
Commercial vehicles (\textbf{a}): $\rho = 0.24$ (non-significant), with the vast majority clamped near SOH\,=\,100\%.
E-GMP~192S (\textbf{b}): $\rho = 0.10$ (non-significant).
Niro/Kona (\textbf{c}): $\rho = 0.17$ (non-significant).
E-GMP~180S (\textbf{d}): $\rho = 0.62$ ($\textit{p} = 0.005$, CC\mbox{-}filtered $n = 19$;
unfiltered $\rho = -0.18$, $n = 63$).
\textbf{e,} Relative capacity distribution for MEB vehicles (no SOH available). The bimodal pattern likely reflects unresolved battery pack variants (different net capacity trims) that cannot be distinguished through OBD-II telemetry metadata.
\textbf{f,} Summary of SOH--capacity correlations across platforms, with E-GMP~180S as best case ($R^2 = 38.3$\%).}
\label{fig:soh}
\end{figure}

The core finding of this study is the dramatic inconsistency of BMS~SOH reliability across model platforms (Fig.~\ref{fig:soh}).
We compare the independently measured capacity gaps (dQ) with the corresponding BMS~SOH differences for each platform.

\textbf{Commercial vehicles (Porter/Bongo~EV, 90S).}
Despite exhibiting the largest actual capacity range (24.7\% dQ gap, $\textit{p} = 0.0003$), the vast majority report BMS~SOH at or near 100\% (mean SOH Q1\,=\,99.6\%, Q4\,=\,100.0\%; Fig.~\ref{fig:soh}a). The SOH indicator is effectively non-functional---a 25\% capacity difference produces only a 0.4 percentage-point SOH gap ($\textit{p} = 0.39$, non-significant), making the true health disparity invisible to the owner, warranty system, and any regulatory assessment.

\textbf{E-GMP (192S).}
Vehicles show a 16.4\% actual capacity gap ($\textit{p} = 0.0002$), yet the BMS~SOH gap is only 1.0 percentage point ($\textit{p} = 0.479$, non-significant). The dQ--SOH correlation is $\rho = 0.10$ ($\textit{p} = 0.534$), indicating that BMS~SOH is essentially random with respect to actual capacity (Fig.~\ref{fig:soh}b).

\textbf{Niro/Kona (98S).}
A 20.7\% actual capacity gap ($\textit{p} = 0.004$) corresponds to a non-significant SOH gap of 1.6 percentage points ($\textit{p} = 0.504$), with $\rho = 0.17$ (Fig.~\ref{fig:soh}c).

\textbf{E-GMP~180S.}
The E-GMP~180S subgroup (original IONIQ5~LR with 180-series cells, separated from 192S due to different pack configuration) is the only Korean platform where BMS~SOH shows moderate correlation with independently measured capacity under strict CC-filtering conditions. The dQ--SOH correlation is $\rho = 0.62$ ($\textit{p} = 0.005$, $n = 19$), and the SOH gap of 4.9 percentage points approaches significance ($\textit{p} = 0.058$). However, this correlation is unstable: in the full unfiltered sample ($n = 63$), the correlation reverses sign ($\rho = -0.18$, non-significant), indicating that the positive association depends on strict sample selection and is not robust (Extended Data Table~\ref{tab:sensitivity}). Moreover, the E-GMP~180S has the smallest inter-vehicle capacity variation among Korean platforms (dQ~CV\,=\,4.6\%, ground truth CV\,=\,1.4\%), which limits the practical significance of this correlation. The SOH gap dramatically understates the 12.2\% actual capacity difference (Fig.~\ref{fig:soh}d). Notably, the E-GMP~180S shows this correlation while the E-GMP (192S) from the same manufacturer does not ($\rho = 0.10$), suggesting that SOH algorithm quality varies even within the same manufacturer.

\textbf{MEB platform (Audi Q4 / VW ID.4, 96S).}
These 382~vehicles---the largest single-platform group in our dataset---do not expose SOH through the on-board diagnostics (OBD-II) telemetry interface, although the vehicles likely compute SOH internally.
The dQ distribution is markedly bimodal (Fig.~\ref{fig:soh}e), with two non-overlapping clusters separated by a $\sim$27 percentage-point gap---likely reflecting distinct battery pack variants (e.g., different net capacity trims) that cannot be resolved from OBD-II telemetry metadata, as all vehicles are registered under the same 96S/82\,kWh specification.
The 10.3\% overall capacity CV thus combines both inter-trim and intra-trim variation, representing a completely hidden health disparity.
Owners, insurers, and regulators have no visibility into the battery condition of these vehicles (Fig.~\ref{fig:soh}e).

The SOH reliability summary (Fig.~\ref{fig:soh}f) makes the inconsistency stark: the correlation between actual capacity and BMS~SOH ranges from non-existent (MEB: SOH not exposed via OBD-II) to noise (E-GMP: $\rho = 0.10$) to moderate in a restricted subset (E-GMP~180S: $\rho = 0.62$ for CC-filtered $n = 19$, but $\rho = -0.18$ for unfiltered $n = 63$). Even E-GMP~180S---the best-performing platform under controlled conditions---achieves $R^2 = 38.3$\% in the CC-filtered subset, meaning BMS~SOH misses over 60\% of real capacity variation. No platform achieves a correlation above 0.7, which would represent minimally adequate tracking.

\subsection*{Usage patterns explain health variance}

Having established that real capacity differences exist and that BMS~SOH fails to capture them, we next investigate what drives these differences.
Using a Weighted Degradation Factor (WDF) approach---adapted from the wear density function\cite{Han2014} and extended with principal component analysis (PCA) dimensionality reduction of binned usage patterns (state-of-charge (SOC) distribution, temperature exposure, C-rate distribution) followed by Ridge regression with within-model pooling to account for model heterogeneity---we find that usage patterns consistently explain substantial health variance beyond mileage alone across all platforms (Extended Data Fig.~1).
Because dQ requires constant-current slow charging sessions available for only a subset of vehicles (65--93\% depending on platform), we use kWh/\%SOC---computable from routine driving data for virtually all vehicles---as the primary health indicator for usage-pattern analysis.
To verify that this metric choice does not introduce artifacts, we confirmed the SOC dose--response directly in dQ: high-SOC-usage vehicles show 2.1--4.6\% lower dQ than low-SOC-usage vehicles for Commercial, E-GMP, and E-GMP~180S platforms (all $\textit{p} < 0.02$; partial $\rho = -0.21$ to $-0.49$ after controlling for mileage), confirming that the efficiency-based findings reflect genuine electrochemical capacity loss.

The hierarchical $R^2$ analysis demonstrates this clearly.
For kWh/\%SOC (driving energy efficiency, our primary cross-manufacturer health indicator), the within-model pooled WDF cross-validated $R^2$ ranges from 0.34 (E-GMP~180S) to 0.85 (Niro/Kona), substantially higher than subgroup-level estimates that conflate inter-model differences with usage effects (Extended Data Fig.~1a, Extended Data Table~\ref{tab:fullwdf}).
Mileage alone explains modest variance for MEB ($+0.28$) and E-GMP ($+0.22$) but negligible or negative variance for E-GMP~180S ($-0.03$), Commercial ($-0.07$), and Niro/Kona ($-0.06$) (Extended Data Fig.~1b).
The partial WDF contribution---variance explained by usage patterns after accounting for mileage---ranges from $+0.16$ (MEB) to $+0.91$ (Niro/Kona), confirming that usage patterns explain 16--91\% of health variance that mileage alone cannot.

\begin{figure}[H]
\centering
\includegraphics[width=\textwidth]{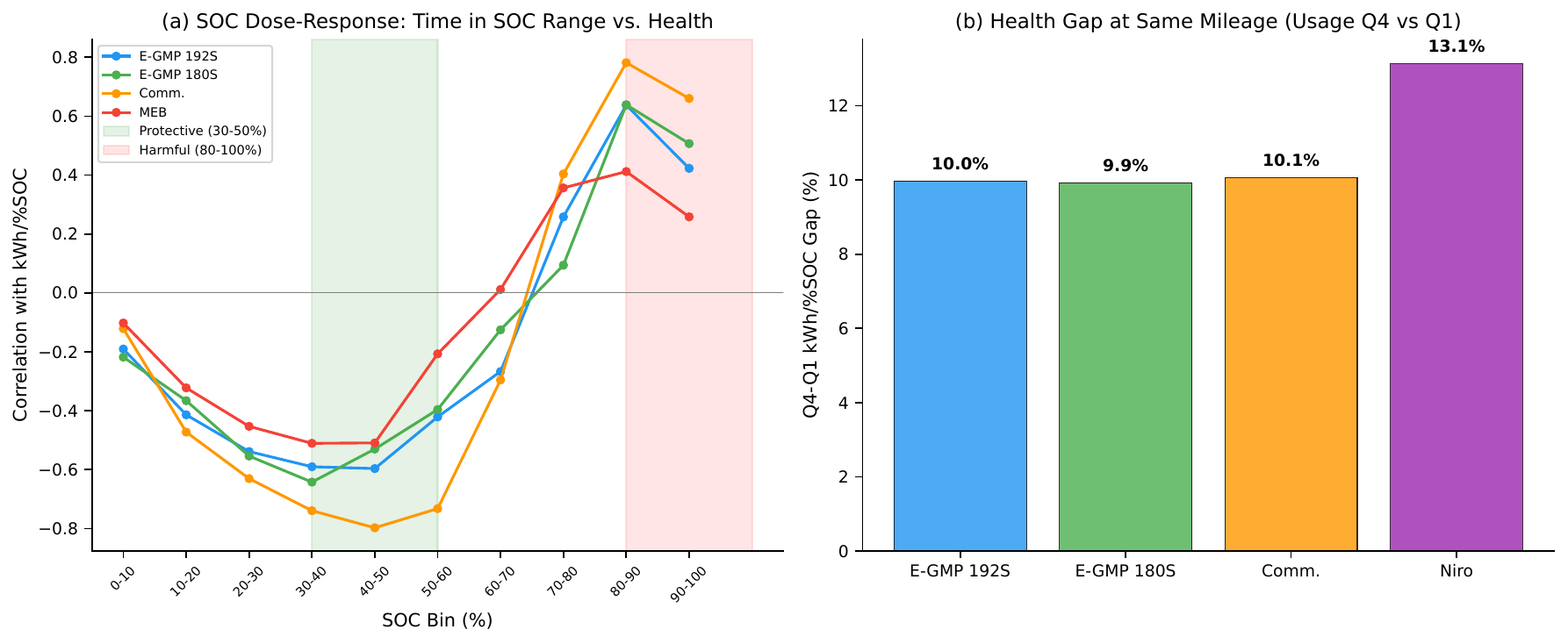}
\caption{\textbf{SOC dose--response association with battery health.}
\textbf{a,} Time fraction in each SOC bin vs.\ health outcome across platforms.
High SOC ($>80\%$) is consistently associated with worse health, while moderate SOC (30--50\%) is associated with better health outcomes.
\textbf{b,} Mileage-stratified comparison showing that at the same mileage, high-SOC-usage vehicles show $\sim$10--13\% worse energy efficiency.}
\label{fig:dose}
\end{figure}

The SOC dose--response analysis reveals the primary usage factor: time spent at high SOC is consistently associated with worse battery health, while time at moderate SOC (30--50\%) is associated with better health outcomes (Fig.~\ref{fig:dose}a).
This pattern replicates across all platforms and is consistent with established electrochemical understanding of nickel-manganese-cobalt (NMC) cathode degradation at high voltage\cite{Keil2016,Schmalstieg2014,Woody2020,Wikner2018}, where calendar aging accelerates with increasing SOC due to elevated electrode potential driving SEI growth and cathode structural degradation\cite{Yang2017,Peterson2010,Marongiu2015}.
At the same mileage band, vehicles in the highest SOC-usage quartile show on average 10--13\% worse energy efficiency than those in the lowest quartile (Fig.~\ref{fig:dose}b).
This efficiency gap corresponds to direct capacity differences: within the same SOC-usage comparison, dQ-based capacity is 2--5\% lower in the highest quartile for three Korean-manufacturer platforms ($\textit{p} < 0.02$), consistent with accelerated calendar aging at elevated electrode potential.
These dose--response associations are consistent with several of Bradford~Hill's guidelines for causation assessment\cite{Hill1965}---including strength of association, consistency across platforms, biological gradient (monotonic SOC--degradation relationship), and plausibility (established NMC electrochemistry)---though observational data cannot exclude unmeasured confounding.

Notably, among the three usage dimensions captured in our WDF framework---SOC distribution, ambient temperature exposure, and C-rate distribution---SOC usage dominates health outcomes.
Ambient temperature shows no significant correlation with health in any platform ($\rho = 0.08$--$0.20$, all $\textit{p} > 0.07$), likely because thermal management systems effectively buffer the battery from external temperature variation in real-world Korean driving conditions.
Similarly, fast-charging exposure (high C-rate fraction) shows weak or inconsistent associations ($\rho = -0.33$ to $+0.03$).
These findings contrast with laboratory expectations where temperature and C-rate are major degradation accelerators\cite{Keil2016,Schmalstieg2014}, suggesting that under real-world conditions with active thermal management, how owners use their SOC window matters more than ambient climate or charging speed.

\subsection*{BMS-independent universal health markers}

Given the unreliability of BMS~SOH, we sought electrochemical indicators that could serve as manufacturer-independent alternatives.
Through systematic exploration of differential capacity (dQ/dV) features extracted from slow charging sessions, we validated the dQ/dV peak voltage position ($V_{\text{peak}}$)---a laboratory-established degradation indicator\cite{Bloom2005,Bloom2005b,Dubarry2012}---as a field-deployable universal health marker (Extended Data Fig.~2a).

\textbf{Physical basis.}
In NMC/graphite cells, lithium inventory loss shifts the dominant dQ/dV peak from 3.9--4.1~V (healthy) to 3.6--3.7~V (degraded), reflecting cathode phase transitions and graphite staging changes\cite{Bloom2005,Bloom2005b,Dubarry2012,Bloom2010,Birkl2017,Lewerenz2017}.
Crucially, this measurement requires only voltage and current data---no open-circuit voltage (OCV)--SOC lookup table, no BMS SOC, and no manufacturer-specific calibration.

\textbf{Predictive power.}
Using within-model pooled Fisher-z correlations, $V_{\text{peak}}$ correlates with kWh/\%SOC across all five NMC platforms ($\rho = 0.60$--$0.72$, all $\textit{p} < 0.001$; Extended Data Fig.~2a).
As a WDF prediction target, $V_{\text{peak}}$ shows complementary predictability to kWh/\%SOC: usage patterns predict $V_{\text{peak}}$ better for E-GMP ($R^2 = 0.39$ vs.\ 0.25 for kWh) but predict kWh/\%SOC better for other platforms (Extended Data Fig.~2b).

\textbf{Universal threshold.}
A $V_{\text{peak}}$ threshold of 3.67~V classifies vehicles as degraded or healthy with 74--89\% accuracy across all platforms via leave-one-subgroup-out (LOSO) validation (Extended Data Fig.~3b).
This threshold was not optimized for any specific platform but emerges consistently from the electrochemical properties of NMC chemistry\cite{Bloom2006,Harlow2019}.

\textbf{Threshold robustness.}
The 3.67~V threshold is robust to voltage quantization differences across platforms (97.5--252.5~mV; LOSO accuracy consistent at 74--89\%), temperature restriction (15--30\textdegree C shifts threshold by $<$0.02~V with $<$2\%p accuracy change), and SOC window selection (slow CC charging operates over gradual $\sim$0.1~V/cell traversals).

\textbf{Augmented WDF.}
Adding $V_{\text{peak}}$ as an additional predictor to the PCA-5 usage bins improves health prediction for platforms where usage patterns alone leave substantial unexplained variance.
For the MEB platform, within-model pooled $R^2$ increases from 0.40 (usage bins only, $n = 288$) to 0.62 (augmented with $V_{\text{peak}}$), a $+0.22$ absolute improvement (Extended Data Fig.~2c).
E-GMP also benefits ($+0.08$; $R^2 = 0.25 \to 0.33$; $n = 157$).
In contrast, platforms where usage patterns already achieve high $R^2$---Commercial (0.76), Niro/Kona (0.77), and E-GMP~180S (0.74)---show no augmentation benefit, as expected when the usage-pattern model already captures most health-relevant variance.

\subsection*{Cross-manufacturer pattern transfer}

A key question is whether degradation patterns learned from one manufacturer can predict health in another.
Using normalized LOSO validation---where each platform's health values are z-scored to remove absolute scale differences---we find that usage-pattern-to-degradation relationships transfer substantially across platforms (Extended Data Fig.~3a).

Normalized LOSO $R^2$ ranges from 0.16 (MEB) to 0.68 (E-GMP~180S) for usage bins alone. Adding dQ features further improves Korean-platform transfer
(E-GMP: 0.40; Commercial: 0.69; E-GMP~180S: 0.72)
but not MEB ($R^2 \leq 0$) or Niro/Kona ($0.35 \to 0.23$), consistent with MEB's architectural isolation and Niro/Kona's already-saturated usage-pattern model.
The pairwise transfer matrix (Extended Data Table~\ref{tab:loso}) reveals that models from the same manufacturer transfer well: E-GMP~192S $\to$ Commercial achieves $R^2 = 0.66$, and E-GMP~192S $\to$ E-GMP~180S achieves $R^2 = 0.63$.
The MEB platform is relatively isolated---no other platform transfers successfully to MEB, likely reflecting its distinct cell architecture (Samsung SDI cells in MEB vs.\ SK Innovation/LG Energy cells in Korean platforms)\cite{Schmuch2018,Baumhofer2014}.

This finding has practical significance: it suggests that a health assessment model trained on Hyundai/Kia data could provide meaningful predictions for other Korean-platform vehicles, but that separate calibration is needed for fundamentally different battery architectures.
More broadly, it demonstrates that the relationship between usage patterns and degradation follows sufficiently universal electrochemical principles\cite{Preger2020,Schmalstieg2014} that cross-manufacturer transfer learning is feasible when appropriate normalization is applied.

\section*{Discussion}

Our results reveal a fundamental problem in the current EV ecosystem: the metric that consumers, insurers, and regulators rely on for battery health assessment---BMS~SOH---is unreliable in a model-dependent and unpredictable way.
The implications are significant.

First, \textbf{warranty assessments based on BMS~SOH are unreliable}.
A commercial vehicle owner whose battery has lost 25\% of its capacity will never trigger an SOH-based warranty claim because the BMS reports 100\%.
An EV6 owner whose battery health has declined more than their neighbor's identical vehicle cannot know this from SOH alone ($\rho = 0.10$).
This represents a structural information asymmetry\cite{Akerlof1970} that benefits the warrantor at the expense of the consumer.

Second, \textbf{the used EV market lacks a reliable health signal}.
Unlike vehicle mileage---which, despite its limitations, provides a standardized and independently verifiable metric---BMS~SOH varies in meaning and availability across manufacturers.
A buyer comparing a used IONIQ5 (E-GMP~180S, where SOH shows moderate correlation with capacity only under strict CC-filtering conditions) with a used VW ID.4 (where SOH is not accessible through the OBD-II interface) cannot make an informed comparison.
This problem is compounded by the growing second-life battery market, where accurate health assessment is critical for safe repurposing\cite{Zhu2021,Martinez2018,Hesse2017}.
The EU Battery Regulation\cite{EU2023} and California's ACC~II requirements\cite{CARB2022} mandate SOH reporting, but our findings demonstrate that the quality of SOH reporting, not merely its availability, requires regulatory attention.

Third, and most surprisingly, \textbf{SOH quality varies within a single manufacturer}.
Hyundai's E-GMP~180S shows $\rho = 0.62$ between dQ and SOH in a CC-filtered subset ($n = 19$), though this reverses to $\rho = -0.18$ in the full sample ($n = 63$); the same company's E-GMP shows near-zero correlation regardless of filtering (unfiltered $\rho = 0.02$, CC-filtered $\rho = 0.10$).
This inconsistency is not merely a cross-manufacturer problem but a within-manufacturer one, further undermining consumer confidence.

Beyond these reliability concerns, the magnitude of hidden health heterogeneity has direct economic consequences.
In commercial fleets where worst-quartile vehicles have 25\% less capacity yet all report SOH\,=\,100\%, the aggregate exposure for large fleets reaches millions of dollars in hidden replacement costs\cite{Zhu2021}---invisible to buyers and unrecoverable through warranty claims.
In the used EV market, within-model capacity variation of 12--25\% creates a valuation gap that, without independent health metrics, drives adverse selection\cite{Akerlof1970}: informed sellers of degraded vehicles benefit while sellers of well-maintained vehicles cannot differentiate their product.
Moreover, our data challenge the conventional reliance on mileage as a proxy for battery condition in used EV transactions: mileage alone explains negligible health variance for most platforms ($R^2 \leq 0.28$), while usage patterns---particularly SOC management---explain 16--91\% of variance beyond mileage (Extended Data Fig.~1b). A used EV buyer who selects solely on low mileage may acquire a vehicle with significantly worse battery health than a higher-mileage alternative driven with more favorable charging habits.

\begin{figure}[H]
\centering
\includegraphics[width=\textwidth]{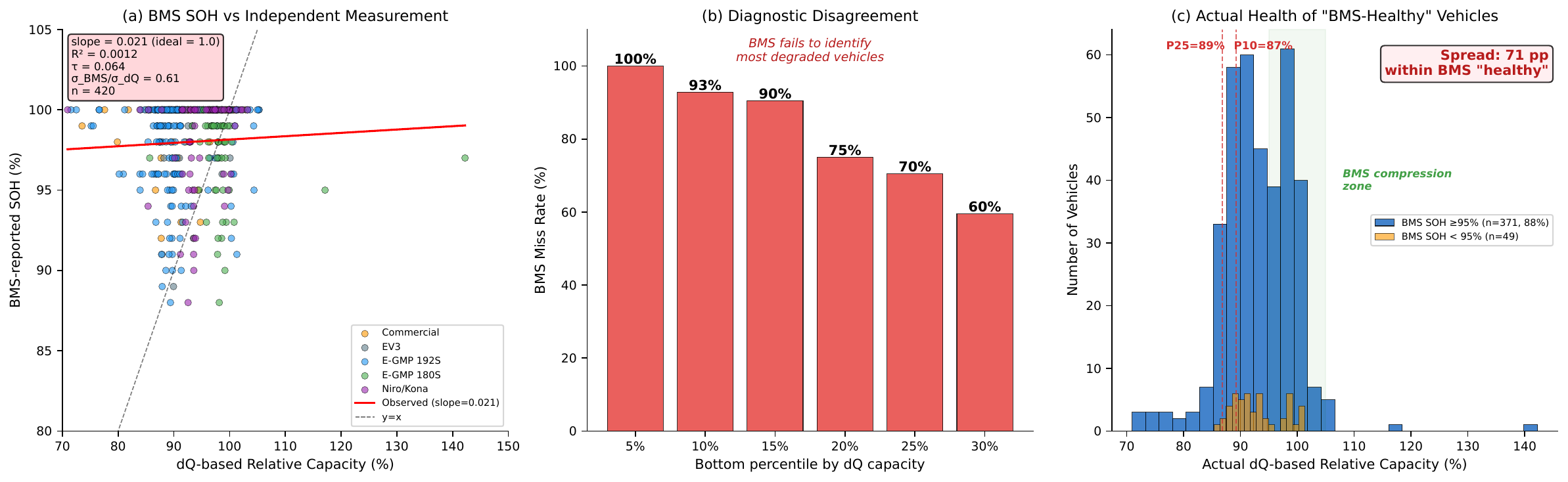}
\caption{\textbf{Warranty misclassification analysis.}
\textbf{a,} BMS-reported SOH vs.\ dQ-based relative capacity ($n = 420$). The regression slope ($+0.02$; ideal $= 1.0$) confirms BMS~SOH is independent of actual capacity. Points color-coded by platform.
\textbf{b,} Diagnostic disagreement rate: fraction of vehicles in each bottom percentile by dQ that BMS fails to identify as worst. BMS misses 100\% of the bottom 5\% and 93\% of the bottom 10\%.
\textbf{c,} Distribution of actual dQ-based capacity for vehicles BMS calls ``healthy'' (SOH $\geq 95$\%). Despite uniform BMS classification, actual capacity spans 71--142\%.}
\label{fig:warranty}
\end{figure}

To quantify the warranty implications directly, we compared BMS~SOH with our independent dQ-based relative capacity for 420~vehicles with both measurements across five platform subgroups that report BMS~SOH (excluding MEB, which does not expose SOH via OBD-II; including EV3) (Fig.~\ref{fig:warranty}a).
The regression slope is $+0.02$ ($R^2 = 0.001$, $\textit{p} = 0.49$), confirming that BMS~SOH is statistically independent of actual capacity.
Among the worst-condition 10\% of vehicles by dQ capacity, 93\% are not identified as worst by BMS---they are invisible to any SOH-based warranty trigger (Fig.~\ref{fig:warranty}b).
Of the 371~vehicles that BMS classifies as ``healthy'' (SOH $\geq 95$\%), actual relative capacity spans 71\% to 142\% (total spread of 71~percentage points) hidden behind a nominally reassuring number (Fig.~\ref{fig:warranty}c).
This diagnostic disagreement means that warranty systems built on BMS~SOH will systematically fail to identify the most degraded vehicles in a fleet.

Regarding measurement limitations, the E-GMP~180S CC-filtered correlation ($\rho = 0.62$, $n = 19$) warrants careful interpretation. The sign reversal in the unfiltered sample ($\rho = -0.18$, $n = 63$) indicates dependence on strict selection criteria, and the platform's minimal capacity variation (dQ~CV\,=\,1.4\%) makes rank correlation inherently unstable (Extended Data Table~\ref{tab:sensitivity}). A post-hoc power analysis confirms that $n = 19$ provides only 59\% power to detect $\rho = 0.5$ (below the 80\% threshold).

Regarding generalizability, all vehicles were monitored in South Korea, where temperate climate conditions and well-developed charging infrastructure may limit extrapolation to extreme-climate regions (e.g., Scandinavia, Middle East). Furthermore, while the fleet spans five manufacturer brands, these belong to two corporate groups (Hyundai Motor Group and Volkswagen Group), and BMS~SOH validation is confined to Hyundai Motor Group platforms since Volkswagen Group vehicles do not expose SOH via OBD-II. The self-selected nature of the fleet---owners who opted into aftermarket monitoring---may introduce selection bias toward battery-conscious consumers.

Despite these caveats, the result is informative: that E-GMP~180S shows some tracking under controlled conditions demonstrates that our dQ methodology captures real differences that BMS~SOH can partially detect, confirming that near-zero correlations in other models reflect genuine BMS deficiencies rather than measurement artifacts. The E-GMP~180S uses a different BMS calibration than the 192-series E-GMP, suggesting that algorithm refinement improved tracking for this subgroup but was not propagated to other models.

Our findings suggest that standardized, manufacturer-independent health assessment is both necessary and feasible.
The $V_{\text{peak}}$ threshold of 3.67~V---which requires only voltage and current data, no proprietary algorithms---achieves 74--89\% classification accuracy across all platforms.
The augmented WDF framework (usage patterns + $V_{\text{peak}}$) achieves within-model pooled $R^2 = 0.33$--$0.76$ across platforms, with the electrochemical feature providing the largest gains where usage patterns alone are least predictive (MEB: $+0.22$; Extended Data Table~\ref{tab:augmented}).
The complementarity is not coincidental: for Commercial and Niro/Kona, usage patterns already explain 75--85\% of health variance, leaving little room for $V_{\text{peak}}$ to contribute additional information.
Notably, usage patterns predict $V_{\text{peak}}$ poorly for Niro/Kona ($R^2 = 0.04$) and modestly for Commercial ($R^2 = 0.24$; Extended Data Fig.~2b), suggesting that $V_{\text{peak}}$ variation in these vehicles reflects initial cell-level manufacturing variability (e.g., different cell suppliers within the same platform\cite{Schmuch2018}) rather than usage-driven electrochemical aging.
In contrast, for MEB and E-GMP, $V_{\text{peak}}$ captures degradation signatures that usage statistics alone cannot resolve.
These results provide a technical foundation for several policy recommendations:

\begin{enumerate}
\item \textbf{Mandatory standardized testing protocols.} Regular capacity tests under standardized conditions (similar to our CC charging protocol) should be required, with results made available to vehicle owners and independent third parties. Benchmark test methodologies already exist in the literature\cite{Harlow2019} and could be adapted for field use.

\item \textbf{BMS~SOH validation requirements.} Manufacturers should be required to demonstrate that their BMS~SOH correlates with independently measured capacity (e.g., $\rho > 0.7$) as a condition for warranty certification\cite{Mohtat2021}. Our data show that no current platform meets this threshold in real fleet conditions.

\item \textbf{Open battery health data standards.} Telematics data sufficient to compute BMS-independent health indicators (voltage, current, temperature during charging) should be available through standardized interfaces, enabling third-party health assessment\cite{Deng2020}. Recent impedance-based approaches\cite{Jones2022} and formation-protocol optimization\cite{Weng2023} demonstrate that battery characterization data can be leveraged for fleet-scale health monitoring.
\end{enumerate}

Several limitations should be acknowledged.
First, our dataset spans approximately 12~months (375~days), which captures early-life degradation patterns but may not fully represent end-of-life behavior where non-linear aging ``knees'' can emerge\cite{Attia2022}.
Moreover, preliminary longitudinal analysis of monthly dQ trajectories showed that seasonal temperature confounding dominated the signal, precluding reliable temporal-sequence evidence from this observation window; longer multi-year datasets will be needed to establish causal temporality.
Second, the dQ measurement has an intra-vehicle CV of 5--7\%, meaning per-vehicle precision is moderate; our findings are strongest for population-level comparisons. However, laboratory validation on a cell of the same type used in E-GMP~180S fleet vehicles (Extended Data Fig.~4) confirms that partial-window dQ tracks RPT capacity with $\rho > 0.80$ even under extreme degradation (SOH declining to 9.7\%), providing experimental support for the proxy's validity.
Third, the multi-window consistency validation shows marginal results for MEB (CV\,=\,5.8\%) and Niro/Kona (CV\,=\,7.1\%), suggesting that measurement precision varies across platforms.
For MEB specifically, the overall capacity CV of 10.3\% combines both inter-trim variation (from the bimodal distribution) and intra-trim degradation variation, which cannot be fully disentangled from OBD-II telemetry data alone.
Fourth, the E-GMP~180S CC-filtered result ($\rho = 0.62$, $n = 19$) suggests that BMS~SOH can show moderate correlation with capacity under controlled conditions, although this correlation is unstable (sign reversal in the full sample; Extended Data Table~\ref{tab:sensitivity}). Our claim is about inconsistency rather than universal failure---a more nuanced but arguably more concerning argument, as consumers cannot know \emph{a priori} which category their vehicle falls into or under which conditions SOH may be informative.
Fifth, while we identify SOC-dependent degradation patterns consistent with Bradford~Hill's guidelines\cite{Hill1965}, we cannot fully separate these from confounds such as driving style, climate, and charger type from observational data alone.

Despite these limitations, the core finding---that BMS~SOH reliability varies dramatically and unpredictably across models---is robust and has immediate practical implications for consumer protection, market transparency, and regulatory oversight.


\clearpage

\begin{table}[ht]
\centering
\caption{Capacity heterogeneity and BMS~SOH comparison under controlled CC charging conditions.}
\label{tab:capacity}
\small
\begin{tabular}{@{}lccccccc@{}}
\toprule
Platform & $n$ & dQ CV (\%) & Q4--Q1 Gap (\%) & Gap $\textit{p}$ & SOH Gap (\%p) & SOH $\textit{p}$ & $\rho$(dQ, SOH) \\
\midrule
E-GMP (192S) & 59 & 6.2 & 16.4 & 0.0002*** & 1.0 & 0.479 & 0.10 \\
E-GMP~180S & 36 & 4.6 & 12.2 & 0.014* & 4.9 & 0.058 & 0.62$^{\dagger\dagger}$ \\
Commercial (90S) & 52 & 8.8 & 24.7 & 0.0003*** & 0.4 & 0.391 & 0.24 \\
Niro/Kona (98S) & 33 & 6.7 & 20.7 & 0.004** & 1.6 & 0.504 & 0.17 \\
MEB (96S) & 152 & 10.3 & --- & --- & --- & N/A & N/A \\
\bottomrule
\end{tabular}
\vspace{2pt}
\noindent\parbox{\textwidth}{\footnotesize *$\textit{p} < 0.05$; **$\textit{p} < 0.01$; ***$\textit{p} < 0.001$ (Gap $\textit{p}$, Mann--Whitney $U$ test). $^{\dagger\dagger}$$\textit{p} < 0.01$ ($\rho$(dQ, SOH), Spearman correlation). MEB vehicles do not expose SOH through OBD-II.}

\noindent\parbox{\textwidth}{\footnotesize $n$ = vehicles with qualifying CC sessions meeting strict criteria (voltage window 3.60--3.72~V/cell, current CV $< 25$\%, temperature 10--35\textdegree C). $\rho$(dQ, SOH) computed on the subset with both CC-filtered dQ and BMS~SOH available ($n = 39$, 19, 28, 23 respectively). Fig.~\ref{fig:soh} scatter plots show all vehicles with any available dQ measurement for visual context; annotated $\rho$ values match this table. Unfiltered correlations (all vehicles with both dQ and SOH, without CC-session filtering) are: Commercial $\rho = 0.43$ ($n = 81$, $\textit{p} < 0.001$), E-GMP $\rho = 0.02$ ($n = 180$), Niro/Kona $\rho = 0.25$ ($n = 61$, $\textit{p} = 0.06$), E-GMP~180S $\rho = -0.18$ ($n = 63$); see Extended Data Table~\ref{tab:sensitivity} for full sensitivity analysis.}
\end{table}
\clearpage

\textbf{Sample size notation.}
We distinguish four sample sizes throughout: $N_{\mathrm{total}}$, the total number of vehicles per platform; $n_{\mathrm{dQ}}$, vehicles with valid CC-filtered dQ measurements; $n_{\mathrm{SOH}}$, vehicles with both dQ and BMS~SOH; and $n_{\mathrm{CC}}$, the subset meeting all rigorous CC-filtering criteria (current CV$<$25\%, temperature 10--35\textdegree C) used for Spearman correlations in Table~\ref{tab:capacity}.
In scatter plots (Fig.~\ref{fig:soh}), $n$ denotes the number of plotted points (after interquartile range (IQR) outlier removal from $n_{\mathrm{SOH}}$) and $n_{\mathrm{CC}}$ denotes the CC-filtered subset from which $\rho$ is computed.

\begin{table}[H]
\centering
\caption{Sensitivity of within-model pooled results to minimum model size threshold.}
\label{tab:sensitivity_minn}
\small
\begin{tabular}{@{}lcccc@{}}
\toprule
& \multicolumn{4}{c}{Minimum model size ($n_{\min}$)} \\
\cmidrule(l){2-5}
Platform & 7 & 10 & 15 & 20 \\
\midrule
\multicolumn{5}{@{}l}{\textit{bins $\to$ kWh/\%SOC $R^2$ (Extended Data Fig.~1a)}} \\
MEB & 0.40 (3m) & 0.40 (3m) & 0.40 (3m) & 0.40 (3m) \\
E-GMP (192S) & 0.63 (7m) & 0.56 (5m) & 0.64 (3m) & 0.64 (3m) \\
Commercial & 0.75 (2m) & 0.75 (2m) & 0.75 (2m) & 0.75 (2m) \\
Niro/Kona & 0.85 (3m) & 0.85 (3m) & 0.85 (3m) & 0.85 (3m) \\
E-GMP~180S & 0.34 (1m) & 0.34 (1m) & 0.34 (1m) & 0.34 (1m) \\[4pt]
\multicolumn{5}{@{}l}{\textit{$V_{\text{peak}}$ $\leftrightarrow$ kWh/\%SOC pooled $\rho$ (Extended Data Fig.~2a)}} \\
MEB & 0.72 (3m) & 0.72 (3m) & 0.74 (2m) & 0.74 (2m) \\
E-GMP (192S) & 0.69 (5m) & 0.69 (3m) & 0.69 (3m) & 0.72 (2m) \\
Commercial & 0.72 (2m) & 0.72 (2m) & 0.72 (2m) & 0.72 (2m) \\
Niro/Kona & 0.60 (2m) & 0.60 (2m) & 0.60 (2m) & 0.42 (1m) \\
E-GMP~180S & 0.68 (1m) & 0.68 (1m) & 0.68 (1m) & 0.68 (1m) \\
\bottomrule
\end{tabular}
\vspace{2pt}
\noindent\parbox{\textwidth}{\footnotesize Number of vehicle models included shown in parentheses (e.g., 3m = 3~models). Results are stable across thresholds; E-GMP shows minor variation ($R^2$: 0.56--0.64) as small models (GV70: $n=7$, IONIQ5~N~NE: $n=7$) are included or excluded. Niro/Kona $\rho$ drops at $n_{\min}=20$ because only one model (Kona, $n=23$) remains.}
\end{table}
\clearpage

\section*{Figure Legends}

\clearpage
\section*{Methods}

\subsection*{Data collection}

Telematics data were collected from 1,114~EVs enrolled in a commercial vehicle monitoring service (Betterwhy Inc.) over approximately 12~months (375~days) (February 2025 to February 2026). The fleet comprises vehicles whose owners subscribed to aftermarket OBD-II telematics monitoring for battery health tracking; the sample is therefore not a random draw from the national EV population and may over-represent owners concerned about battery condition. All vehicle owners provided informed consent for de-identified data to be used for research and development purposes. The study uses only de-identified, aggregated vehicle-level statistics and was determined to be exempt from institutional review board approval under Kyungpook National University's research ethics guidelines.
Vehicles span five manufacturers: Hyundai ($n = 369$), Kia ($n = 333$), Genesis ($n = 25$), Audi ($n = 195$), and Volkswagen ($n = 186$), plus 3~vehicles from other manufacturers (SsangYong, GM) and 3~with mixed manufacturer labels (assigned to their respective platform groups in Extended Data Table~\ref{tab:fleet}).
Data include pack voltage, current, SOC, temperature, system accumulated capacity (SAC), individual cell voltages, mileage, and BMS state flags at approximately 1-second resolution.
Vehicle-to-model mapping was obtained from the vehicle information database table.
Data were stored in a TimescaleDB instance with hypertable partitioning for efficient time-range queries.

All data were collected via an OBD-II telematics device installed in each vehicle, transmitting to a cloud monitoring platform operated by the fleet management company.
This data pathway is representative of the third-party monitoring ecosystem: OBD-II ports provide standardized access to selected BMS parameters, but manufacturers retain discretion over which internal calculations are exposed through this interface.
MEB-platform vehicles (Audi, Volkswagen) compute battery health internally---as required for their warranty assessment systems---but do not expose SOH values through the OBD-II telemetry channel accessible to third-party devices.
This constitutes a concrete example of the information asymmetry we identify: the manufacturer possesses health information that is structurally withheld from the vehicle owner and independent monitoring systems.
Of the 1,114~vehicles, 730 (65.5\%) have BMS-reported SOH values; the remaining 384---including all 382 Audi/VW MEB-platform vehicles---do not expose SOH through the OBD-II telemetry interface used by third-party monitoring platforms.

\subsection*{Usage profile extraction}

For each vehicle, we computed usage profiles consisting of:
(a) 10~SOC bins (0--10\%, 10--20\%, \ldots, 90--100\%) representing the fraction of time spent in each SOC range;
(b) 5~temperature bins ($< 5$\textdegree C, 5--15\textdegree C, 15--25\textdegree C, 25--35\textdegree C, $> 35$\textdegree C);
(c) 5~C-rate bins based on per-vehicle nominal ampere-hours; and
(d) aggregate statistics (total mileage, mean SOC, mean temperature, data duration).
Nominal ampere-hours were computed as $\text{nominal\_kWh} \times 1000 / (\text{cell\_count} \times 3.7\,\text{V})$.

\subsection*{Health indicator extraction}

Five independent health indicators were computed per vehicle, drawing on established methods for in-field battery diagnostics\cite{Birkl2017,Smith2017}:

\begin{enumerate}
\item \textbf{kWh/\%SOC}: Median energy consumption per SOC percentage during driving episodes ($|\text{current}| > 5$~A, duration $> 5$~min, $\Delta\text{SOC} > 5\%$).
Note that BMS~SOC estimation (instantaneous charge level, typically $\pm$2--3\% accuracy via voltage-based and coulomb-counting methods\cite{Plett2004,Xiong2018}) is a fundamentally simpler and better-established measurement than BMS~SOH (long-term capacity fade estimation requiring degradation modeling).
As a within-model relative health indicator, any systematic SOC bias cancels in cross-vehicle comparisons; we independently confirmed that kWh/\%SOC-based findings replicate in the BMS-independent dQ metric (see Results).

\item \textbf{DCIR}: Median DC internal resistance computed from current-step transients during driving ($\Delta I > 30$~A, $\Delta t < 5$~s), standardized to 25\textdegree C using Arrhenius correction ($E_a = 20$~kJ/mol)\cite{Richter2017}.

\item \textbf{dQ}: Total accumulated ampere-hours during constant-current slow charging sessions within a standardized voltage window (3.60--3.72~V/cell, manufacturer-dependent), with edge trimming and current/temperature controls.

\item \textbf{Thermal impedance}: Temperature rise per integrated $I^2 dt$ during fast charging sessions (current $> 50$~A, duration $> 10$~min)\cite{Lin2014}.

\item \textbf{Cell voltage standard deviation}: Standard deviation of individual cell voltages under load ($|\text{current}| > 10$~A), reflecting cell imbalance that grows with aging\cite{Baumhofer2014}.
\end{enumerate}

\subsection*{CC charging session identification}

Charging sessions were identified using the BMS \texttt{chg\_state\,=\,1} flag.
To validate this flag's reliability across manufacturers, we developed a physics-only CC segment detector using a state machine approach: current between 3--50~A with CV\,$< 25\%$, monotonically increasing voltage, increasing SAC, and duration $> 5$~minutes.
Comparison against the flag-based method across 16~sample vehicles from 8~model groups yielded Precision\,=\,88\% and Recall\,=\,91\%.
All physics-only segments had \texttt{chg\_state\,=\,1} (100\% concordance).
Flag-only segments that the physics detector missed were exclusively rapid charging events ($> 50$~A) intentionally excluded from the physics criteria.

\subsection*{Rigorous dQ measurement protocol}

For the SOH comparison analysis, we applied strict controls:
voltage windows of 0.08--0.12~V/cell (auto-selected per subgroup for maximum coverage), with boundaries aligned to pack voltage measurement quantization steps;
session current CV\,$< 25\%$;
temperature 10--35\textdegree C;
edge trimming of first and last 3~minutes;
minimum 2~qualifying sessions per vehicle.
Pack voltage quantization was empirically detected per platform: E-GMP\,=\,97.5~mV, MEB\,=\,252.5~mV, Commercial/Niro\,=\,102.5~mV.

For visualization and cross-vehicle comparison, raw dQ values are expressed as relative capacity (\%) by normalizing to the within-platform 90th percentile (P90): $\text{Relative capacity} = \text{dQ} / \text{P90}_{\text{platform}} \times 100$.
P90 was chosen as a robust near-new reference, less sensitive to outliers than the maximum; consequently, approximately 10\% of vehicles in each platform exhibit values exceeding 100\%, indicating capacity above this reference rather than a measurement artifact.

\subsection*{Multi-window consistency validation}

For each vehicle pair with dQ measurements in $\geq 3$ common voltage windows (4 non-overlapping windows of 0.1~V/cell each), we computed the dQ ratio per window and then the coefficient of variation of these ratios.
If dQ faithfully represents total capacity, the ratio should be window-independent (CV~$\to$~0).
Analysis was limited to 2,000 randomly sampled pairs per subgroup for computational feasibility.

\subsection*{Ground truth validation}

Vehicles with wide-range CC charging sessions ($\geq 0.25$~V/cell) were identified to serve as in-field analogues of laboratory capacity tests.
For each qualifying vehicle, both wide-window (0.25~V/cell) and narrow-window (0.10~V/cell) dQ values were computed using fixed voltage ranges (not session-dependent SOC ranges) to avoid confounding.
Spearman rank correlation and Pearson correlation (log-transformed dQ) between the two windows assessed ranking agreement.

\subsection*{Lab cell validation}

To provide experimental ground truth for the dQ proxy methodology, we analyzed accelerated aging data from an SK Innovation 55.6~Ah NCM pouch cell---the same cell type used in IONIQ~5 (E-GMP~180S) vehicles in our fleet dataset.
The cell was cycled at 10~A (0.18C) CC-CV charging (2.5--4.2~V) with periodic RPT measuring full-range capacity as the ground truth.
Over 198 cycles, SOH declined from 100\% to 9.7\%, providing a wide dynamic range for validation.

For each cycle, we extracted dQ (accumulated charge) across 60 voltage windows with lower bounds ranging from 2.5 to 3.5~V and upper bounds from 3.3 to 4.2~V, selected to ensure sufficient voltage span for meaningful charge accumulation.
Charging segments were identified using current-based detection ($|I| > 100$~mA) rather than voltage-derivative detection, as the high-resolution data (1~s sampling) produces numerous zero-$\Delta V$ steps that cause artifacts in voltage-based detection.
For each window, we computed Spearman rank correlation ($\rho$) and mean absolute error (MAE) of the inter-cycle dQ ratio relative to the RPT capacity ratio, analogous to the multi-window consistency validation applied to the fleet data.

\subsection*{Weighted Degradation Factor (WDF)}

Building on the wear density function concept originally proposed for battery depreciation modeling\cite{Han2014}, we adapt the WDF framework to fleet-scale health analysis.
Usage profiles (SOC bins, temperature bins, C-rate bins, and optionally dQ/dV features) were reduced to five principal components (PCA-5\cite{Jolliffe2002}; Extended Data Fig.~1c) and used in Ridge regression\cite{HoerlKennard1970} ($\alpha = 1.0$) to predict health indicators.
To account for model heterogeneity within each platform (e.g., different battery specifications or cell suppliers), we employ a within-model pooled approach: leave-one-out cross-validation (LOOCV) is performed within each vehicle model, and predictions from all models within a platform are pooled to compute a single cross-validated $R^2$.
This prevents Simpson's paradox that can arise when mixing models with different health baselines.
Models with fewer than 7~vehicles (or fewer than $n_{\mathrm{PCA}}+2$) were excluded; a sensitivity analysis varying this threshold from 7 to 20 confirms that all main results are robust (Table~\ref{tab:sensitivity_minn}).
For cross-platform correlation analysis, within-model Spearman $\rho$ values were pooled via Fisher z-transformation with inverse-variance weighting ($w_i = n_i - 3$) and back-transformed to obtain a pooled $\rho$\cite{Fisher1921}.
Hierarchical $R^2$ analysis decomposed the total explained variance into mileage-only, WDF-only, and partial WDF (variance explained by usage beyond mileage).

\subsection*{Augmented WDF}

The augmented WDF framework extends the standard PCA-5 usage-bin features with the dQ/dV peak voltage ($V_{\text{peak}}$) as an additional predictor, preserving it as a separate feature rather than including it in the PCA decomposition.
This design preserves the electrochemical information in $V_{\text{peak}}$ while maintaining the dimensionality reduction of usage patterns.
The same within-model pooled LOOCV approach is used for both bins-only and augmented models to ensure fair comparison.

\subsection*{Transfer learning (Normalized LOSO)}

Leave-one-subgroup-out validation was performed by training on all subgroups except one and testing on the held-out subgroup.
To enable transfer despite different absolute health scales across platforms, both features and targets were z-score normalized within each subgroup (zero mean, unit variance).
Pairwise transfer matrices were computed by training on one source subgroup and testing on each target subgroup individually\cite{Dechent2021}.

\subsection*{Statistical analysis}

All statistical tests are two-sided. Exact $\textit{p}$-values are reported throughout; where $\textit{p} < 0.001$, we report $\textit{p} < 0.001$. Significance levels: $*$ $\textit{p} < 0.05$, $**$ $\textit{p} < 0.01$, $***$ $\textit{p} < 0.001$. Exact sample sizes ($n$) for each analysis are provided in the corresponding figure legends and tables.

All analyses were performed in Python~3.11 using NumPy, SciPy, scikit-learn, and pandas. The following statistical methods were used: Spearman rank correlation (SOH--capacity associations), Pearson correlation (WDF model evaluation), Fisher $z$-transformation (confidence intervals for correlation coefficients), bootstrap resampling ($N = 1{,}000$ iterations; 95\% confidence intervals reported as error bars and bracketed ranges), and the Mann--Whitney $U$ test\cite{MannWhitney1947} (quartile comparisons of health indicators).

Bonferroni correction is applied to the four platform-level SOH correlation tests (adjusted $\alpha = 0.0125$); the E-GMP~180S CC-filtered correlation ($\rho = 0.62$, $n = 19$, $\textit{p} = 0.005$) remains significant after correction. However, this significance does not extend to the unfiltered sample ($\rho = -0.18$, $n = 63$; Extended Data Table~\ref{tab:sensitivity}).

\subsection*{Data availability}

Vehicle-level de-identified summary statistics underlying all figures and tables are deposited at Zenodo (DOI: 10.5281/zenodo.18896326). Raw telematics data contain proprietary vehicle identifiers and driving patterns that preclude public release; de-identified session-level data are available from the corresponding author upon reasonable request, subject to a data use agreement that protects vehicle owner privacy. Source Data for all main figures are provided with this paper.

\subsection*{Code availability}

All analysis code used to generate the results in this study is available at \url{https://github.com/SekyungHan/battery-epidemiology} and archived at Zenodo (DOI: 10.5281/zenodo.18896326).

\subsection*{Acknowledgements}

The authors thank Betterwhy Inc. for developing and operating the large-scale vehicle telematics monitoring platform that enabled this study, and for supporting the multi-manufacturer fleet enrolment and long-term data collection infrastructure. This research received no external funding.

\subsection*{Author Contributions}

J.P. performed the cross-platform SOH validation analysis, developed the electrochemical marker classification pipeline, conducted the dose-response modelling, and drafted the figures. S.H. conceived and supervised the study, designed the overall analytical framework, and wrote the manuscript. K.K. built and maintained the vehicle telematics data pipeline and performed data quality assurance. S.G. contributed to fleet enrolment and vehicle metadata curation. J.L. and H.S. assisted with data preprocessing and statistical analysis. All authors reviewed and approved the final manuscript.

\subsection*{Competing Interests}

S.H., K.K., and S.G. are affiliated with Betterwhy Inc., which operates the vehicle telematics monitoring platform used to collect the data analyzed in this study. The commercial affiliation did not influence the study design, data analysis, or interpretation of results.

\subsection*{Additional Information}

Reprints and permissions information is available at www.nature.com/reprints.

Correspondence and requests for materials should be addressed to S.H. (skhan@knu.ac.kr).

\setcounter{table}{0}
\renewcommand{\tablename}{Extended Data Table}

\section*{Extended Data}

\subsection*{Extended Data Figure 1}
\noindent\includegraphics[width=\textwidth]{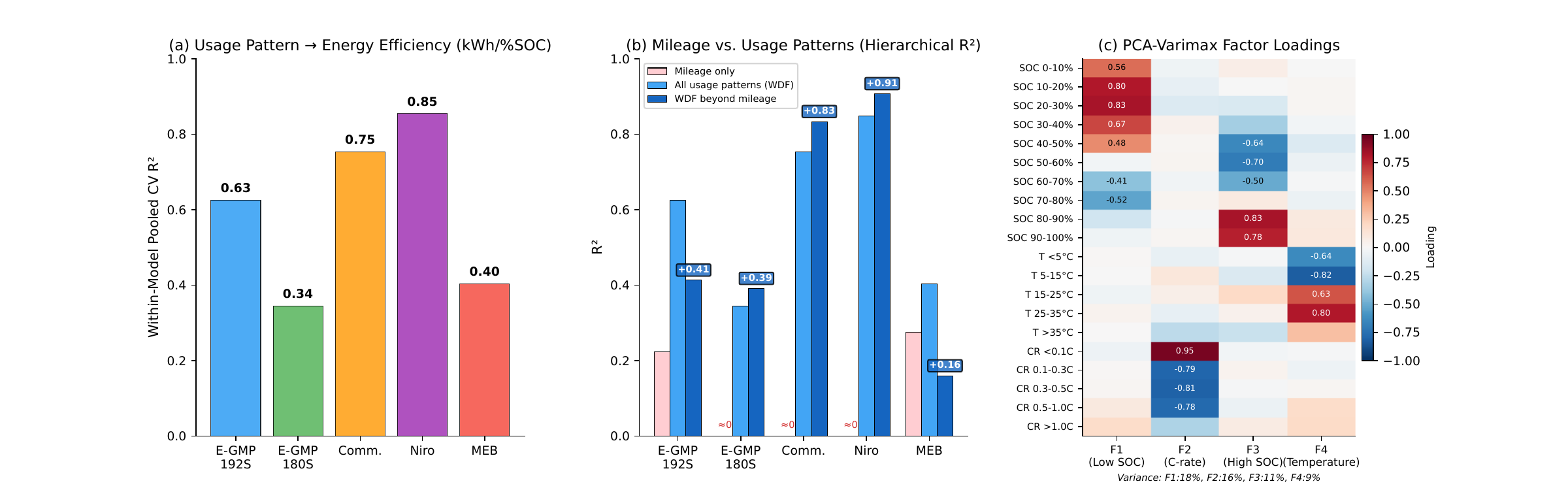}\\
\textbf{Usage patterns explain health variance beyond mileage.}
\textbf{a,} Per-platform within-model pooled cross-validated $R^2$ between usage-pattern WDF predictions and observed energy efficiency (kWh/\%SOC).
\textbf{b,} Hierarchical $R^2$ decomposition showing mileage-only, WDF-only, and partial WDF contributions to kWh/\%SOC prediction. Partial WDF contribution (usage beyond mileage) ranges from $+0.16$ (MEB) to $+0.91$ (Niro/Kona).
\textbf{c,} PCA-Varimax factor loadings identifying SOC-dominated, temperature-dominated, and C-rate-dominated usage factors from 20 binned features.

\subsection*{Extended Data Figure 2}
\noindent\includegraphics[width=\textwidth]{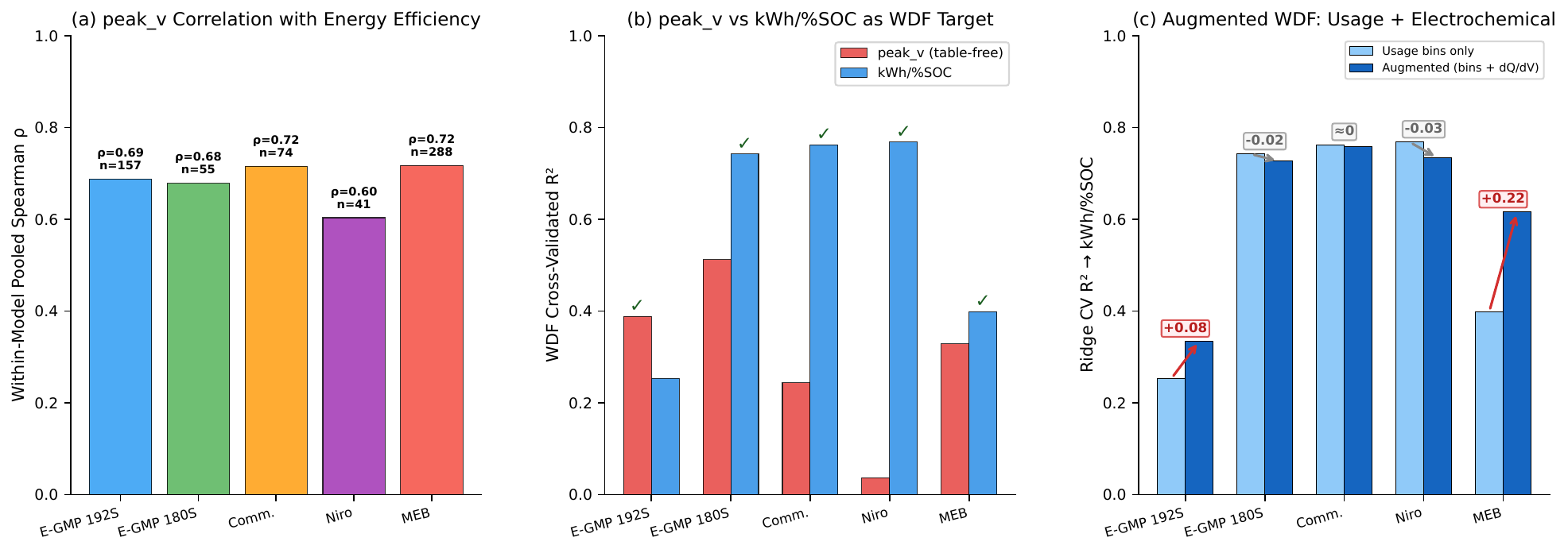}\\
\textbf{dQ/dV peak voltage as a universal, BMS-independent health marker.}
\textbf{a,} Within-model pooled Spearman $\rho$ between $V_{\text{peak}}$ and kWh/\%SOC. All five NMC platforms show significant positive correlation ($\rho = 0.60$--$0.72$, all $\textit{p} < 0.001$).
\textbf{b,} Comparison of $V_{\text{peak}}$ and kWh/\%SOC as WDF targets (checkmark indicates higher within-model pooled $R^2$). $V_{\text{peak}}$ outperforms kWh/\%SOC for E-GMP; kWh/\%SOC outperforms for other platforms.
\textbf{c,} Augmented WDF: adding $V_{\text{peak}}$ to usage bins improves $R^2$ for MEB ($+0.22$) and E-GMP ($+0.08$), but not for platforms where bins alone already achieve high $R^2$.

\subsection*{Extended Data Figure 3}
\noindent\includegraphics[width=\textwidth]{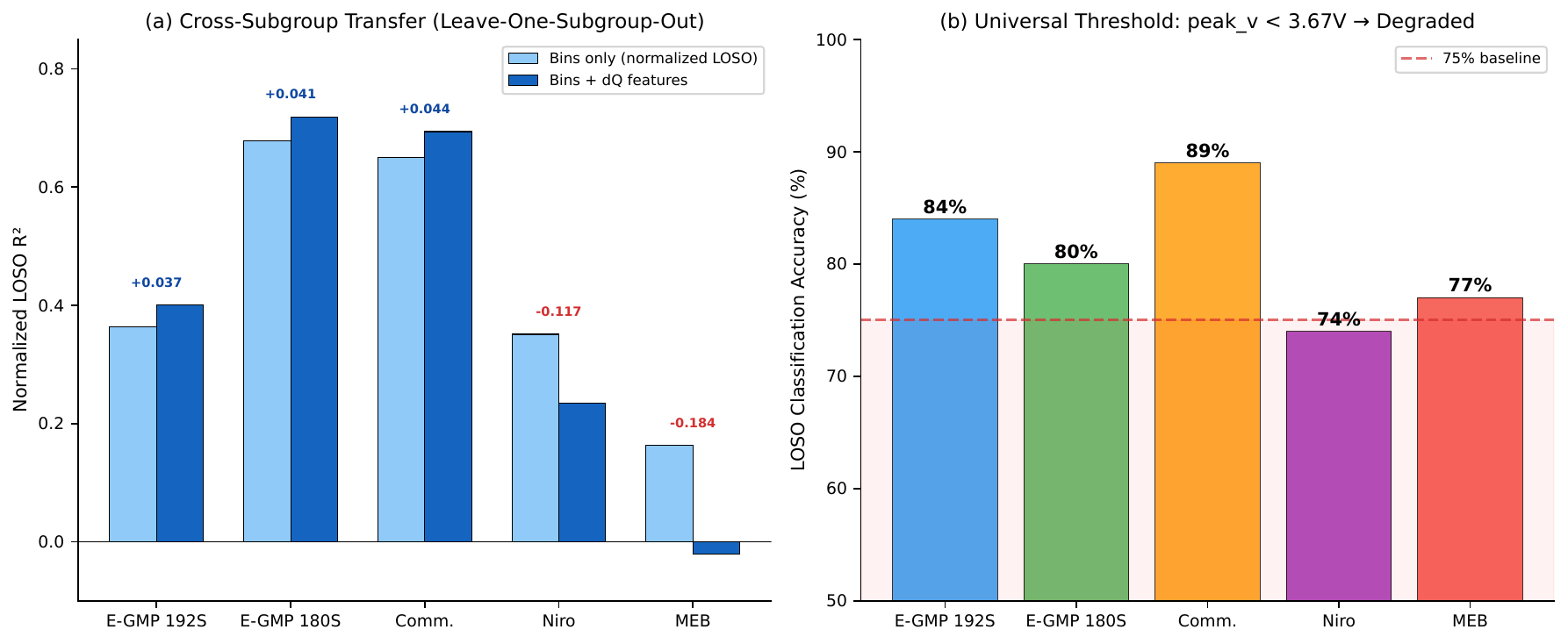}\\
\textbf{Cross-manufacturer transfer learning validation.}
\textbf{a,} Normalized LOSO $R^2$ by platform showing that degradation patterns transfer across manufacturers (0.16--0.68). Adding dQ features generally improves Korean-platform transfer (except Niro/Kona) but not MEB.
\textbf{b,} Universal $V_{\text{peak}}$ threshold of 3.67~V achieves 74--89\% classification accuracy across all NMC platforms via LOSO validation. See Extended Data Table~\ref{tab:loso} for the pairwise transfer matrix.

\subsection*{Extended Data Figure 4}
\noindent\includegraphics[width=\textwidth]{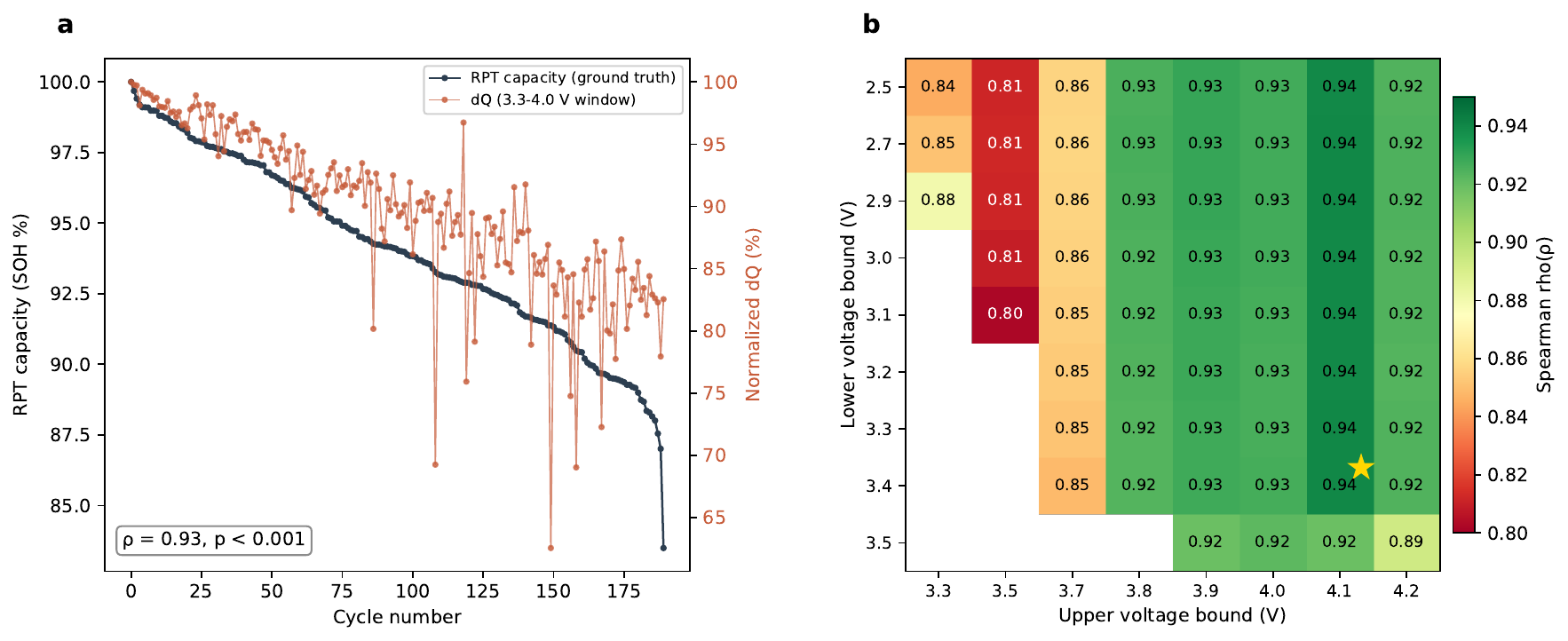}\\
\textbf{Laboratory validation of the dQ proxy using a cell identical to those in E-GMP~180S fleet vehicles.}
An SK Innovation 55.6~Ah NCM pouch cell (identical to those in IONIQ~5 fleet vehicles) was subjected to 198 accelerated aging cycles at 0.18C CC-CV (2.5--4.2~V), with SOH declining from 100\% to 9.7\%.
\textbf{a,} Degradation trajectory showing RPT capacity (ground truth, left axis) and dQ accumulated in the 3.3--4.0~V voltage window (right axis) over 198 cycles. Both metrics decline in parallel, confirming that partial-window dQ faithfully tracks full-range capacity loss.
\textbf{b,} Spearman rank correlation ($\rho$) between partial-window dQ and full RPT capacity across 60 voltage window combinations (lower bound $\times$ upper bound). All windows achieve $\rho > 0.80$ ($\textit{p} < 0.001$); optimal performance ($\rho > 0.93$) clusters at upper bounds $\geq 3.8$~V. Star marks the best window (3.4--4.1~V, $\rho = 0.94$). The lower voltage bound has minimal impact on correlation quality, consistent with the field observation that most capacity-relevant electrochemical activity occurs at higher voltages in NMC cells.

\begin{table}[H]
\centering
\caption{Ground truth validation results. Spearman $\rho$ and Pearson $r$ (log-transformed) between narrow-window (0.10~V/cell) and wide-window ($\geq 0.25$~V/cell) dQ rankings.}
\label{tab:groundtruth}
\small
\begin{tabular}{@{}lccccc@{}}
\toprule
Platform & $n$ & Spearman $\rho$ & $\textit{p}$-value & Pearson $r$ (log) & Conclusion \\
\midrule
E-GMP (192S) & 27 & \textbf{0.913} & $< 0.001$ & 0.966 & PASS \\
Commercial (90S) & 24 & \textbf{0.752} & $< 0.001$ & 0.740 & PASS \\
Niro/Kona (98S) & 23 & 0.523 & 0.010 & \textbf{0.956} & MARGINAL$^\dagger$ \\
E-GMP~180S & 10 & 0.382 & 0.276 & 0.197 & FAIL$^*$ \\
\bottomrule
\multicolumn{6}{@{}l@{}}{$^\dagger$Log-linear relationship strong despite moderate rank correlation.} \\
\multicolumn{6}{@{}l@{}}{$^*$Minimal inter-vehicle variation (dQ CV\,=\,1.4\%) renders rank correlation unstable.}
\end{tabular}
\end{table}

\begin{table}[H]
\centering
\caption{Sensitivity analysis: unfiltered vs.\ CC-filtered BMS~SOH correlations with independent capacity (dQ). Unfiltered correlations use all vehicles with both dQ and BMS~SOH from the merged vehicle-level dataset; CC-filtered correlations use vehicles meeting strict constant-current charging criteria (Table~\ref{tab:capacity}). Unfiltered 95\% CIs are from bootstrap resampling (1,000 iterations); CC-filtered CIs are Fisher $z$-transform analytical estimates. Ground truth dQ~CV is from wide-window ($\geq 0.25$~V/cell) charging sessions (Extended Data Table~\ref{tab:groundtruth}). E-GMP~180S's sign reversal between CC-filtered and unfiltered samples highlights the instability of correlations when inter-vehicle capacity variation is small (CV\,=\,1.4\%). For Commercial vehicles, the unfiltered correlation ($\rho = 0.43$, $n = 81$) exceeds the CC-filtered value ($\rho = 0.24$, $n = 28$); this likely reflects statistical power loss from the three-fold sample size reduction rather than a meaningful difference, as the confidence intervals overlap substantially.}
\label{tab:sensitivity}
\small
\begin{tabular}{@{}lccccccc@{}}
\toprule
\multirow{2}{*}{Platform} & \multicolumn{3}{c}{Unfiltered} & \multicolumn{3}{c}{CC-filtered} & Ground truth dQ \\
\cmidrule(lr){2-4} \cmidrule(lr){5-7}
 & $n$ & $\rho$ & 95\% CI & $n$ & $\rho$ & 95\% CI & CV (\%) \\
\midrule
Commercial (90S) & 81 & 0.43*** & [0.25, 0.56] & 28 & 0.24 & [$-$0.14, 0.56] & 3.6 \\
E-GMP (192S) & 180 & 0.02 & [$-$0.12, 0.17] & 39 & 0.10 & [$-$0.22, 0.41] & 7.1 \\
Niro/Kona (98S) & 61 & 0.25$^\dagger$ & [0.01, 0.47] & 23 & 0.17 & [$-$0.26, 0.54] & 9.6 \\
E-GMP~180S & 63 & $-$0.18 & [$-$0.42, 0.06] & 19 & 0.62** & [0.23, 0.84] & 1.4 \\
MEB (96S) & --- & --- & --- & --- & --- & --- & --- \\
\bottomrule
\end{tabular}
\vspace{0.5em}

\noindent\parbox{\textwidth}{\footnotesize ***$\textit{p} < 0.001$; **$\textit{p} < 0.01$; $^\dagger$$\textit{p} < 0.1$. MEB vehicles do not expose SOH through OBD-II.}
\end{table}

\begin{table}[H]
\centering
\caption{Full subgroup WDF analysis results.}
\label{tab:fullwdf}
\small
\begin{tabular}{@{}lcccccc@{}}
\toprule
Subgroup & $n$ & kWh WDF $R^2$ & Partial WDF & dQ CV (\%) & DCIR CV (\%) & SOH \\
\midrule
E-GMP (192S) & 226 & 0.63 & +0.41 & 6.2 & 15.7 & $\rho=0.10$ \\
E-GMP~180S & 81 & 0.34 & +0.39 & 4.6 & 16.4 & $\rho=0.62$**$^\dagger$ \\
Commercial & 100 & 0.75 & +0.83 & 8.8 & 7.4 & $\rho=0.24$ \\
Niro/Kona & 87 & 0.85 & +0.91 & 6.7 & 7.6 & $\rho=0.17$ \\
MEB & 379 & 0.40 & +0.16 & 10.3 & 11.2 & \textbf{N/A} \\
\bottomrule
\end{tabular}
\vspace{0.5em}
\noindent\parbox{\textwidth}{\footnotesize Within-model pooled LOOCV $R^2$. $n$ = vehicles used in analysis (models with $\geq 7$ vehicles). $^\dagger$CC-filtered subset only ($n = 19$); unfiltered $\rho = -0.18$ ($n = 63$). See Extended Data Table~\ref{tab:sensitivity}.}
\end{table}

\begin{table}[H]
\centering
\caption{Pairwise cross-platform transfer matrix ($R^2$, WDF with dQ features). $\leq\!0$ indicates negative transfer ($R^2 \leq 0$).}
\label{tab:loso}
\small
\begin{tabular}{@{}lccccc@{}}
\toprule
Train $\backslash$ Test & MEB & E-GMP~192S & Commercial & Niro & E-GMP~180S \\
\midrule
MEB & --- & $\leq$0 & $\leq$0 & $\leq$0 & $\leq$0 \\
E-GMP~192S & $\leq$0 & --- & 0.66 & $\leq$0 & 0.63 \\
Commercial & $\leq$0 & 0.08 & --- & 0.02 & 0.42 \\
Niro & $\leq$0 & 0.10 & 0.47 & --- & 0.33 \\
E-GMP~180S & $\leq$0 & 0.24 & 0.09 & $\leq$0 & --- \\
\bottomrule
\end{tabular}
\end{table}

\begin{table}[H]
\centering
\caption{Augmented WDF improvement by platform.}
\label{tab:augmented}
\small
\begin{tabular}{@{}lcccc@{}}
\toprule
Platform & $n$ & Bins $R^2$ & Aug.\ $R^2$ & $\Delta R^2$ \\
\midrule
MEB (96S) & 288 & 0.40 & 0.62 & \textbf{+0.22} \\
E-GMP (192S) & 157 & 0.25 & 0.33 & \textbf{+0.08} \\
Commercial (90S) & 74 & 0.76 & 0.76 & $<$0.01 \\
Niro/Kona (98S) & 41 & 0.77 & 0.73 & $-$0.03 \\
E-GMP~180S & 55 & 0.74 & 0.73 & $-$0.02 \\
\bottomrule
\end{tabular}
\vspace{4pt}

\noindent\parbox{\textwidth}{\footnotesize $n$ = vehicles with $V_{\text{peak}}$ and kWh/\%SOC available. Within-model pooled LOOCV $R^2$. Augmented model adds $V_{\text{peak}}$ as a separate feature to PCA-5 usage bins.}
\end{table}

\begin{table}[H]
\centering
\caption{Complete vehicle fleet composition.}
\label{tab:fleet}
\small
\begin{tabular}{@{}llcccc@{}}
\toprule
Manufacturer & Platform & Cells & $n$ & SOH & Chemistry \\
\midrule
Hyundai/Kia/Genesis & E-GMP~192S (EV6 / IONIQ6 / GV60/70) & 192 & 232 & Yes & NMC \\
Hyundai/Kia & E-GMP~180S (IONIQ5 LR) & 180 & 85 & Yes & NMC \\
Hyundai/Kia & Niro / Kona & 98 & 94 & Yes & NMC \\
Hyundai/Kia & Porter / Bongo (Commercial) & 90 & 100 & Yes & NMC \\
Hyundai/Kia & EV3 & 93 & 42 & Yes & NMC \\
Hyundai/Kia & Casper & 84 & 24 & Yes & NMC \\
Hyundai/Kia/Genesis & Other HK (incl.\ EV9) & Various & 152 & Mostly & NMC \\
Audi/VW & Q4/ID.4/ID.5 (MEB) & 96 & 382 & \textbf{No} & NMC \\
Other & SsangYong / GM & Various & 3 & Mixed & Various \\
\midrule
\textbf{Total} & & & \textbf{1,114} & \textbf{730 (65.5\%)} & \\
\bottomrule
\end{tabular}
\end{table}


\begin{thebibliography}{58}

\bibitem{IEA2025} International Energy Agency \textit{Global EV Outlook 2025: Expanding Sales in Diverse Markets} (2025); \url{https://www.iea.org/reports/global-ev-outlook-2025}.

\bibitem{BloombergNEF2025} BloombergNEF \textit{Electric Vehicle Outlook 2025} (2025); \url{https://about.bnef.com/insights/clean-transport/electric-vehicle-outlook/}.

\bibitem{Schmuch2018} Schmuch, R., Wagner, R., H{\"o}rpel, G., Placke, T. \& Winter, M. Performance and cost of materials for lithium-based rechargeable automotive batteries. \textit{Nat. Energy} \textbf{3}, 267--278 (2018).

\bibitem{Wassiliadis2022} Wassiliadis, N. et al. Quantifying the state of the art of electric powertrains in battery electric vehicles: Range, efficiency, and lifetime from component to system level of the Volkswagen ID.3. \textit{eTransportation} \textbf{12}, 100167 (2022).

\bibitem{Tian2020} Tian, H., Qin, P., Li, K. \& Zhao, Z. A review of the state of health for lithium-ion batteries: Research status and suggestions. \textit{J. Clean. Prod.} \textbf{261}, 120813 (2020).

\bibitem{Berecibar2016} Berecibar, M. et al. Critical review of state of health estimation methods of Li-ion batteries for real applications. \textit{Renew. Sustain. Energy Rev.} \textbf{56}, 572--587 (2016).

\bibitem{Li2019} Li, Y. et al. Data-driven health estimation and lifetime prediction of lithium-ion batteries: A review. \textit{Renew. Sustain. Energy Rev.} \textbf{113}, 109254 (2019).

\bibitem{Akerlof1970} Akerlof, G.A. The market for ``lemons'': Quality uncertainty and the market mechanism. \textit{Q. J. Econ.} \textbf{84}, 488--500 (1970).

\bibitem{EU2023} European Parliament and Council \textit{Regulation (EU) 2023/1542 concerning batteries and waste batteries} (Official Journal of the European Union, L 191/1, 2023).

\bibitem{CARB2022} California Air Resources Board \textit{Advanced Clean Cars II: Section 1962.8 Warranty Requirements for Zero-Emission and Batteries in Plug-in Hybrid Electric 2026 and Subsequent Model Year Passenger Cars} (2022); \url{https://ww2.arb.ca.gov/}.

\bibitem{Zhu2021} Zhu, J. et al. End-of-life or second-life options for retired electric vehicle batteries. \textit{Cell Rep. Phys. Sci.} \textbf{2}, 100537 (2021).

\bibitem{Martinez2018} Martinez-Laserna, E. et al. Battery second life: Hype, hope or reality? A critical review of the state of the art. \textit{Renew. Sustain. Energy Rev.} \textbf{93}, 701--718 (2018).

\bibitem{Birkl2017} Birkl, C.R., Roberts, M.R., McTurk, E., Bruce, P.G. \& Howey, D.A. Degradation diagnostics for lithium ion cells. \textit{J. Power Sources} \textbf{341}, 373--386 (2017).

\bibitem{Dubarry2012} Dubarry, M., Truchot, C. \& Liaw, B.Y. Synthesize battery degradation modes via a diagnostic and prognostic model. \textit{J. Power Sources} \textbf{219}, 204--216 (2012).

\bibitem{Baumhofer2014} Baumh{\"o}fer, T., Br{\"u}hl, M., Rothgang, S. \& Sauer, D.U. Production caused variation in capacity aging trend and correlation to initial cell performance. \textit{J. Power Sources} \textbf{247}, 332--338 (2014).

\bibitem{Attia2022} Attia, P.M. et al. ``Knees'' in lithium-ion battery aging trajectories. \textit{J. Electrochem. Soc.} \textbf{169}, 060517 (2022).

\bibitem{Geotab2024} Geotab \textit{EV Battery Degradation: Key Findings from 22,700 Vehicle Data Analysis} (2024); \url{https://www.geotab.com/blog/ev-battery-health/}.

\bibitem{Geotab2025} Geotab \textit{EV Battery Health Study: New Data on Fast Charging and Degradation} (2025); \url{https://www.geotab.com/press-release/ev-battery-health-degradation-fast-charging-study/}.

\bibitem{Recurrent2024} Recurrent Auto \textit{How Long Do Electric Car Batteries Last?} (2024); \url{https://www.recurrentauto.com/research/how-long-do-ev-batteries-last}.

\bibitem{Severson2019} Severson, K.A. et al. Data-driven prediction of battery cycle life before capacity degradation. \textit{Nat. Energy} \textbf{4}, 383--391 (2019).

\bibitem{Attia2020} Attia, P.M. et al. Closed-loop optimization of fast-charging protocols for batteries with machine learning. \textit{Nature} \textbf{578}, 397--402 (2020).

\bibitem{Ng2020} Ng, M., Zhao, J., Yan, Q., Conduit, G.J. \& Seh, Z.W. Predicting the state of charge and health of batteries using data-driven machine learning. \textit{Nat. Mach. Intell.} \textbf{2}, 161--170 (2020).

\bibitem{Roman2021} Roman, D., Saxena, S., Robu, V., Pecht, M. \& Flynn, D. Machine learning pipeline for battery state-of-health estimation. \textit{Nat. Mach. Intell.} \textbf{3}, 447--456 (2021).

\bibitem{Wei2025} Wei, Y., Yao, J., Liu, Y., Bao, M. \& Liu, X. Multi-modal framework for battery state of health evaluation using open-source electric vehicle data. \textit{Nat. Commun.} \textbf{16}, 1052 (2025).

\bibitem{Bloom2005} Bloom, I. et al. Differential voltage analyses of high-power lithium-ion cells: 1. Technique and application. \textit{J. Power Sources} \textbf{139}, 295--303 (2005).

\bibitem{Weng2014} Weng, C., Sun, J. \& Peng, H. A unified open-circuit-voltage model of lithium-ion batteries for state-of-charge estimation and state-of-health monitoring. \textit{J. Power Sources} \textbf{258}, 228--237 (2014).

\bibitem{Pelletier2017} Pelletier, S., Jabali, O., Laporte, G. \& Veneroni, M. Battery degradation and behaviour for electric vehicles: Review and numerical analyses of several models. \textit{Transp. Res. B} \textbf{103}, 158--187 (2017).

\bibitem{DePalma2023} De Palma, A., Gazzarri, J. \& Lee, J. Estimate long-term impact on battery degradation by considering electric vehicle real-world end-use factors. \textit{J. Power Sources} \textbf{573}, 233128 (2023).

\bibitem{Han2014} Han, S., Han, S. \& Aki, H. A practical battery wear model for electric vehicle charging applications. \textit{Appl. Energy} \textbf{113}, 1100--1108 (2014).

\bibitem{Keil2016} Keil, P. et al. Calendar aging of lithium-ion batteries. \textit{J. Electrochem. Soc.} \textbf{163}, A1872--A1880 (2016).

\bibitem{Schmalstieg2014} Schmalstieg, J., K{\"a}bitz, S., Ecker, M. \& Sauer, D.U. A holistic aging model for {Li(NiMnCo)O$_2$} based 18650 lithium-ion batteries. \textit{J. Power Sources} \textbf{257}, 325--334 (2014).

\bibitem{Woody2020} Woody, M., Arbabzadeh, M., Lewis, G.M., Keoleian, G.A. \& Stefanopoulou, A.G. Strategies to limit degradation and maximize Li-ion battery service lifetime: Critical review and guidance for stakeholders. \textit{J. Energy Storage} \textbf{28}, 101231 (2020).

\bibitem{Wikner2018} Wikner, E. \& Thiringer, T. Extending battery lifetime by avoiding high SOC. \textit{Appl. Sci.} \textbf{8}, 1825 (2018).

\bibitem{Yang2017} Yang, X., Leng, Y., Zhang, G., Ge, S. \& Wang, C. Modeling of lithium plating induced aging of lithium-ion batteries: Transition from linear to nonlinear aging. \textit{J. Power Sources} \textbf{360}, 28--40 (2017).

\bibitem{Peterson2010} Peterson, S.B., Apt, J. \& Whitacre, J. Lithium-ion battery cell degradation resulting from realistic vehicle and vehicle-to-grid utilization. \textit{J. Power Sources} \textbf{195}, 2385--2392 (2010).

\bibitem{Marongiu2015} Marongiu, A., Roscher, M. \& Sauer, D.U. Influence of the vehicle-to-grid strategy on the aging behavior of lithium battery electric vehicles. \textit{Appl. Energy} \textbf{137}, 899--912 (2015).

\bibitem{Hill1965} Hill, A.B. The environment and disease: Association or causation? \textit{Proc. R. Soc. Med.} \textbf{58}, 295--300 (1965).

\bibitem{Bloom2005b} Bloom, I. et al. Differential voltage analyses of high-power lithium-ion cells: 2. Applications. \textit{J. Power Sources} \textbf{139}, 304--313 (2005).

\bibitem{Bloom2010} Bloom, I. et al. Differential voltage analyses of high-power lithium-ion cells: 4. Cells containing NMC. \textit{J. Power Sources} \textbf{195}, 877--882 (2010).

\bibitem{Lewerenz2017} Lewerenz, M., Marongiu, A., Warnecke, A. \& Sauer, D.U. Differential voltage analysis as a tool for analyzing inhomogeneous aging: A case study for LiFePO$_4$|graphite cylindrical cells. \textit{J. Power Sources} \textbf{368}, 57--67 (2017).

\bibitem{Bloom2006} Bloom, I., Christophersen, J., Abraham, D. \& Gering, K. Differential voltage analyses of high-power lithium-ion cells: 3. Another anode phenomenon. \textit{J. Power Sources} \textbf{157}, 537--542 (2006).

\bibitem{Harlow2019} Harlow, J.E. et al. A wide range of testing results on an excellent lithium-ion cell chemistry to be used as benchmarks for new battery technologies. \textit{J. Electrochem. Soc.} \textbf{166}, A3031--A3044 (2019).

\bibitem{Preger2020} Preger, Y. et al. Degradation of commercial lithium-ion cells as a function of chemistry and cycling conditions. \textit{J. Electrochem. Soc.} \textbf{167}, 120532 (2020).

\bibitem{Hesse2017} Hesse, H.C., Schimpe, M., Kucevic, D. \& Jossen, A. Lithium-ion battery storage for the grid---A review of stationary battery storage system design tailored for applications in modern power grids. \textit{Energies} \textbf{10}, 2107 (2017).

\bibitem{Mohtat2021} Mohtat, P., Lee, S., Siegel, J.B. \& Stefanopoulou, A.G. Towards better estimability of electrode-level lithium-ion battery parameters from cell-level measurements. \textit{J. Electrochem. Soc.} \textbf{168}, 070502 (2021).

\bibitem{Deng2020} Deng, J., Bae, C., Denlinger, A. \& Miller, T. Electric vehicles batteries: requirements and challenges. \textit{Joule} \textbf{4}, 511--515 (2020).

\bibitem{Jones2022} Jones, P.K., Stimming, U. \& Lee, A.A. Impedance-based forecasting of lithium-ion battery performance amid uneven usage. \textit{Nat. Commun.} \textbf{13}, 4806 (2022).

\bibitem{Weng2023} Weng, A. et al. Predicting the impact of formation protocols on battery lifetime immediately after manufacturing. \textit{Joule} \textbf{7}, 1745--1761 (2023).

\bibitem{Smith2017} Smith, K. et al. Life prediction model for grid-connected Li-ion battery energy storage system. \textit{IEEE Trans. Ind. Appl.} \textbf{53}, 752--762 (2017).

\bibitem{Plett2004} Plett, G.L. Extended Kalman filtering for battery management systems of LiPB-based HEV battery packs: Part 3. State and parameter estimation. \textit{J. Power Sources} \textbf{134}, 277--292 (2004).

\bibitem{Xiong2018} Xiong, R., Cao, J., Yu, Q., He, H. \& Sun, F. Critical review on the battery state of charge estimation methods for electric vehicles. \textit{IEEE Access} \textbf{6}, 1832--1843 (2018).

\bibitem{Richter2017} Richter, F., Kjelstrup, S., Vie, P.J. \& Burheim, O.S. Thermal conductivity and internal temperature profiles of Li-ion secondary batteries. \textit{J. Power Sources} \textbf{359}, 592--600 (2017).

\bibitem{Lin2014} Lin, X. et al. A lumped-parameter electro-thermal model for cylindrical batteries. \textit{J. Power Sources} \textbf{257}, 1--11 (2014).

\bibitem{Jolliffe2002} Jolliffe, I. \textit{Principal Component Analysis}, 2nd edn. (Springer, 2002).

\bibitem{HoerlKennard1970} Hoerl, A.E. \& Kennard, R.W. Ridge regression: Biased estimation for nonorthogonal problems. \textit{Technometrics} \textbf{12}, 55--67 (1970).

\bibitem{Fisher1921} Fisher, R.A. On the ``probable error'' of a coefficient of correlation deduced from a small sample. \textit{Metron} \textbf{1}, 3--32 (1921).

\bibitem{Dechent2021} Dechent, P. et al. Estimation of Li-ion degradation and application to electric vehicle fleets. \textit{iScience} \textbf{24}, 102060 (2021).

\bibitem{MannWhitney1947} Mann, H. \& Whitney, D. On a test of whether one of two random variables is stochastically larger than the other. \textit{Ann. Math. Stat.} \textbf{18}, 50--60 (1947).

\end{thebibliography}
\end{document}